\newtheorem{theorem}{Theorem}
\newtheorem{proposition}{Proposition}
\newtheorem{assumption}{Assumption}
\newtheorem{definition}{Definition}
\newtheorem{lemma}{Lemma}

\newtheorem{conjecture}{Conjecture}

\documentclass[12pt]{article}

\usepackage{amsmath}
\usepackage{amssymb}
\usepackage{amsfonts}

\newcommand{\qed}{{\mbox{} \hspace*{\fill}{\vrule height5pt width4pt
depth0pt}}\\}

\def\M{\hspace*{0.75em}}

\begin{document}

\title{A Link between Sequential Semi-anonymous Nonatomic Games and their Large Finite Counterparts}
\author{Jian Yang\\
Department of Management Science and Information Systems\\
Business School, Rutgers University\\
Newark, NJ 07102\\
Email: jyang@business.rutgers.edu}

\date{July 2015; revised, March 2016; accepted, May 2016}
\maketitle

\begin{abstract}
We show that obtainable equilibria of a multi-period nonatomic game can be used by players in its large finite counterparts to achieve near-equilibrium payoffs. Such equilibria in the form of random state-to-action rules are parsimonious in form and easy to execute, as they are both oblivious of past history and blind to other players' present states. Our transient results can be extended to a stationary case, where the finite multi-period games are special discounted stochastic games. In both nonatomic and finite games, players' states influence their payoffs along with actions they take; also, the random evolution of one particular player's state is driven by all players' states as well as actions. The finite games can model diverse situations such as dynamic price competition. But they are notoriously difficult to analyze. Our results thus suggest ways to tackle these problems approximately.\\


\noindent{\bf Keywords: }Nonatomic Game; Convergence in Probability; $\epsilon$-Equilibrium

\noindent{\bf MSC2000 Classification Codes: }60G50; 91A10; 91A13


\end{abstract}

\newpage

\section{Introduction}\label{introduction}

We show that an equilibrium of a random multi-period game involving a continuum of players can be used to achieve asymptotically equilibrium results for its large finite counterparts. The latter finite games can model competitive situations involving random and action-dependent evolution of players' states which in turn influence period-wise payoffs. Their complex natures make equilibria difficult to locate. In contrast, those for the former continuum-player game are simple in form and relatively easy to obtain. Therefore, a bridge between the two types of games can have broad practical implications.

The former continuum-player game can be termed more formally as a {\em sequential semi-anonymous nonatomic game} (SSNG). In it, a continuum of players interact with one another in multiple periods; also, each player's one-time payoff and random one-period state transition are both swayed by his own state and action, as well as the joint distribution of other players' states and actions. This is indeed the anonymous sequential game studied by Jovanovic and Rosenthal \cite{JR88}. We use the name SSNG just to be consistent with the single-period nonatomic-game (NG) literature, where anonymity has been reserved for a more special case. An SSNG's finite counterpart is almost the same except that only a finite number of players are involved. This more realistic situation is much more difficult to handle.

In a few steps, we demonstrate the usefulness of an SSNG equilibrium in finite multi-period games. 
First, in a precise language, Theorem~\ref{T-converge} describes
the gradual retreat of randomness in finite games as the number $n$ of players tends to $+\infty$. This paves way for Theorem~\ref{main}, which states that an SSNG's conditional equilibria, in terms of random state-to-action rules, can be used by its large finite counterparts to reach asymptotically equilibrium payoffs on average. A further refinement of this result is achieved in Theorem~\ref{main-f}. The above transient results can be extended to the stationary case involving discounted payoffs and infinite horizons; see Theorem~\ref{main-s}. The conditional equilibria that facilitate our study are similar to well-understood distributional equilibria. Their existence is also directly verifiable. 

One practical situation to which our results can be applied concerns dynamic price competition. Here, players may be firms producing one identical product type, states may be combinations of the firms' inventory levels and other static or dynamic characteristics such as unit costs, and actions may be unit prices the firms charge for the product. In every period, the random demand arriving to a firm is dependent on not only its own price but also prices charged by other competitors. The actual sales is further constrained by available inventory. So the player's one-time payoff is a function of both its own state (inventory and probably also cost) and action (price) and of the distribution of others' actions (prices). Moreover, the firm's next-period inventory  level depends on its current level, the random demand, and potentially an exogenously given production schedule. So the random single-period state transition is potentially a function of the same factors involved in the payoff.

It is a difficult task to predict or prescribe what inventory-dependent prices the firms will or should charge over a finite time horizon. This can be further complicated by diverse scenarios where firms have different degrees of knowledge on their competitors' inventory levels and/or costs. Our results, on the other hand, will reveal that the nonatomic counterpart SSNG is easier to tackle. Its equilibria can be plugged back to the actual finite-player situations, without regard to the particularities of the scenarios, and still make reasonably well predictions/prescriptions when the number of players is large enough. We are not equipped to answer how large is ``large enough''. But computational study done in a related pricing setting hinted that player numbers ``in the tens'' seem large enough; see, Yang and Xia \cite{YX13}.

In the remainder of the paper, Section~\ref{literature} surveys the relevant literature. We then spend Sections~\ref{overture} and~\ref{finite-game} on essentials of SSNGs and finite games, respectively. In Section~\ref{convergence}, we demonstrate the key result that state evolutions in large finite games will not veer too far away from their NG counterparts. Section~\ref{equilibria} is devoted to the main transient result and Section~\ref{interpretation} its detailed interpretation. This result is extended to the stationary case in Section~\ref{stationary}. Implications of these results and existence of our kind of equilibria for SSNGs are shown in Section~\ref{existence}. We conclude the paper in Section~\ref{discussion}.

\section{Literature Survey}\label{literature}

From early on, NGs have been used as easier-to-analyze proxies of real finite-player situations, such as in the study of perfect competition.
Systematic research on NG started with Schmeidler \cite{S73}. He formulated a single-period semi-anonymous NG, wherein the joint distribution of other players' identities and actions may affect any given player's payoff. When the action space is finite, Schmeidler established the existence of pure equilibria when the game becomes anonymous, so that other players' influence on a game's outcome is channeled through the marginal action distribution alone.
Mas-Colell \cite{M84} showed the existence of distributional equilibria in anonymous NGs with compact metric action spaces. The latter result was extended by Khan and Sun \cite{KS90} to a case where players
differ on how their preferences over actions are influenced by external action distributions. A survey of related works up until the early 2000s was provided by Khan and Sun \cite{KS02}.

Much attention has been paid to the topic of pure-equilibrium existence. Khan and Sun \cite{KS95} developed a purification scheme involving a countable compact metric action space. Khan and Sun \cite{KS99} used non-standard measures on identity spaces and generalized Schmeidler's pure-equilibrium existence result for more general action spaces. Balder \cite{B02} established pure- and mixed-equilibrium existence results that may be regarded as generalizations of Schmeidler's corresponding results. Other notable works still include Yu and Zhang \cite{YZ07} and Balder \cite{B08}. On the other hand, Khan, Rath, and Sun \cite{KRS99} identified a certain limit to which Schmeidler's result can be extended.
Recently, Khan et al. \cite{KRSY13} took players' diverse bio-social traits into consideration and pinpointed saturation of the player-identity distribution as the key to existence of pure equilibria.

Links between NGs and their finite counterparts were covered in Green \cite{G84}, Housman \cite{H88}, Carmona \cite{C04}, Kalai \cite{K04}, Al-Najjar \cite{A08}, and Yang \cite{Y11}. For multi-period games without changing states, Green \cite{G80}, Sabourian \cite{S90}, and Al-Najjar and Smorodinsky \cite{AS01} showed that equilibria for large games are nearly myopic.

SSNGs are both challenging and rewarding to analyze because in them, very realistically, individual states are subject to sways of players' own actions as well as their opponents' states and actions. Jovanovic and Rosenthal \cite{JR88} established the existence of distributional equilibria for such games. This result was generalized by Bergin and Bernhardt \cite{BB95} to cases involving aggregate shocks. In SSNGs' finite-player counterparts, however, randomness in state evolution will not go away. Besides, a player's ability to observe other players' states and actions might also affect his decision. Presented with these difficulties, it is not surprising that known results on sequential finite-player games are restricted to the stationary setting, where they appear as discounted stochastic games first introduced by Shapley \cite{S53}.
According to Mertens and Parthasrathy \cite{MP87}, Duffie, Geanakoplos, Mas-Colell, and McLennan \cite{DGMM94}, and Solan \cite{S98}, for instance, equilibria known to exist for these games come in quite complicated forms that for real implementation, demand a high degree of coordination among players.

It is therefore natural to ask whether sequential finite-player games can be approximated by their NG counterparts. This question has so far been answered by two unpublished articles. For a case unconcerned with the copula between marginal state and action distributions, Bodoh-Creed \cite{BC12} provided an affirmative answer, and went on to show for certain cases that limits of large-game equilibria of a myopic form, when in existence, are NG equilibria. Also, Yang \cite{Y12} verified for the approximability when both state transitions and action plans are driven by exogenously generated idiosyncratic shocks. Our current study attempts with the most general possible setting, without unduly restricting on ways in which a player's payoff can be influenced by other players' states and actions or ways in which the game can evolve randomly. To achieve results of the same spirit, we have to overcome technical challenges posed by the new phenomenon of sampling from non-product joint probabilities.



Some authors went on to pursue stationary equilibria (SE), which stressed the long-run steady-state nature of individual action plans and system-wide multi-states; see, e.g., Hopenhayn \cite{H92} and Adlakha and Johari \cite{AJ10}.
The oblivious equilibrium (OE) concept as proposed by Weintraub, Benkard, and van Roy \cite{WBvR08}, in order to account for impacts of large players, took the same stationary approach by letting participants beware of only long-run average system states.
Weintraub, Benkard, and van Roy \cite{WBvR11} showed links between equilibria of infinite-player games and their finite-player brethren for a setting where the long-run average system state could be defined. Though applicable to many situations, we caution that the implicit stationarity of SE or OE is incompatible with applications that are transient by nature; for instance, the dynamic pricing game studied by Yang \cite{Y12}.

\section{The Nonatomic Game}\label{overture}

The SSNG is a game in which a continuum of players interact with one another over multiple periods. A realistic and yet complicating feature is that players possess individual states which influence their payoffs along with all players' actions. The random evolutions of these states, meanwhile, are affected by players' actions. Furthermore, the semi-anonymous nature of the game means that not only what was done, but also who did what to the extent at which  states partially reveal player identities, figure large in both payoff formation and state evolution. We now provide a detailed  account of the game.

\subsection{Game Primitives}

For some natural number $\bar t\in\mathbb{N}$, we let periods $1,2,....,\bar t$ serve as regular periods and period $\bar t+1$ as the terminal period. For all periods, we let players' individual states and actions form, respectively, separable metric spaces $S$ and $X$. We further require that both spaces be {\em discrete}. In this paper, such a space always stands for a separable metric space with countably many elements and the additional feature that the minimum of the distances between any two points remains strictly positive. The discreteness requirement will be useful at one occasion. But most of our derivations will work if the spaces were merely separable metric.
Given any separable metric space $A$, we use ${\cal B}(A)$ for its Borel $\sigma$-field and ${\cal P}(A)$ for the set of all probability measures on the measurable space $(A,{\cal B}(A))$.

To each player, other players' states and actions are immediately felt in a semi-anonymous fashion, so that what really matters is the joint distribution of other players' states and actions. This distribution, which we dub ``in-action environment'', is a member of the joint state-action distribution space ${\cal P}(S\times X)$.
In any period $t=1,2,...,\bar t$, a player's state $s\in S$, his action $x\in X$, and the in-action environment $\tau\in {\cal P}(S\times X)$ he faces, together determine his payoff in that period. In particular, there is a function
\begin{equation}\label{d-payoff}
\tilde f_t:S\times X\times {\cal P}(S\times X)\rightarrow [-\bar f_t,\bar f_t],
\end{equation}
where $\bar f_t$ is some positive constant on the real line $\mathbb{R}$. It is required that $\tilde f_t(\cdot,\cdot,\tau)$ be a measurable map from $S\times X$ to $[-\bar f_t,\bar f_t]$ for every $\tau\in {\cal P}(S\times X)$.
For the terminal period $\bar t+1$, we let the payoff be 0 in all circumstances. 

Now we describe individual players' random state transitions. Given separable metric spaces $A$ and $B$, we use ${\cal K}(A,B)$ to represent the space of all kernels from $A$ to $B$. Each member $\kappa\in {\cal K}(A,B)\subseteq ({\cal P}(B))^A$ satisfies that\\
\indent\M (i) $\kappa(a)$ is a member of ${\cal P}(B)$ for each $a\in A$, and \\
\indent\M (ii) for each $B'\in {\cal B}(B)$, the real-valued function $\kappa(\cdot|B')$ is measurable. \\
Note that we have used $\kappa(a|B')$ rather than the more conventional $\kappa(B'|a)$ to denote the conditional probability for $B'\in {\cal B}(B)$ when given $a\in A$. The current notation allows us to always read a formula from left to right.
Now in period $1,2,...,\bar t$, let there be a function
\begin{equation}\label{d-transition}
\tilde g_t:S\times X\times {\cal P}(S\times X)\rightarrow {\cal P}(S),
\end{equation}
so that $\tilde g_t(\cdot,\cdot,\tau)$ is a member of ${\cal K}(S\times X,S)$ for each $\tau\in {\cal P}(S\times X)$. For convenience, we use ${\cal G}(S,X)$ to denote the space of all such functions, or what we shall call ``state transition kernels''. In period $t$, when a player is in individual state $s\in S$, takes action $x\in X$, and faces in-action environment $\tau\in {\cal P}(S\times X)$, there will be a $\tilde g_t(s,x,\tau|S')$ chance for his state in period $t+1$ to be in any $S'\in {\cal B}(S)$.

This setup is versatile enough to embrace different player characteristics. For instance, each $s\in S$ may comprise two components $\theta$ and $\omega$, with the $\tilde g_t$'s defined through~(\ref{d-transition}) dictating that $\theta$ stays static over time to serve as a player's innate type. Certainly, the $\tilde f_t$'s defined through~(\ref{d-payoff}) can have all kinds of trends over $\theta$ to reflect players' varying payoff structures.

\subsection{Evolution of the Environments}

In any period $1,2,...,\bar t,\bar t+1$, by ``pre-action environment'' we mean the state distribution $\sigma\in {\cal P}(S)$ of all players. With $\bar t$, $S$, $X$, $(\tilde f_t|t=1,2,...,\bar t)$, and $(\tilde g_t|t=1,2,...,\bar t)$ all given in the background, we use $\Gamma(\sigma_1)$ to denote an (SS)NG with $\sigma_1\in {\cal P}(S)$ as its initial period-1 pre-action environment. For this NG, we can use $\chi_{[1\bar t]}=(\chi_t\mid t=1,...,\bar t)\in ({\cal K}(S,X))^{\bar t}$ to denote a policy profile. Here, each $\chi_t\in {\cal K}(S,X)$ is a map from a player's state to the player's random action choice. Together with the given initial environment $\sigma_1$, this policy profile will help to generate a deterministic pre-action environment trajectory $\sigma_{[1,\bar t+1]}=(\sigma_t\mid t=1,2,...,\bar t,\bar t+1)\in ({\cal P}(S))^{\bar t+1}$ in an iterative fashion. This process is also intertwined with the formation of in-action environments $\tau_1,\tau_2,...,\tau_{\bar t}$ faced by all players in periods $1,2,...,\bar t$.

More notation is needed to precisely describe this evolution. Given distribution $p\in {\cal P}(A)$ and kernel $\kappa\in {\cal K}(A,B)$ for separable metric spaces $A$ and $B$, there is a natural product $p\otimes \kappa\in {\cal P}(A\times B)$, such that
\begin{equation}\label{s-to-mu}
(p\otimes \kappa)(A'\times B')=\int_{A'}p(da)\cdot \kappa(a|B'),\hspace*{.5in}\forall A'\in {\cal B}(A),B'\in {\cal B}(B).
\end{equation}
Here, $p\otimes \kappa$ is essentially the joint distribution generated by the marginal $p$ and conditional distribution $\kappa$. Obviously, $(p\otimes\kappa)|_A$, the marginal of $p\otimes\kappa$ on $A$, is $p$. At the same time, we use $p\odot \kappa$ to denote the marginal $(p\otimes \kappa)|_B$, which satisfies
\begin{equation}
(p\odot\kappa)(B')=(p\otimes \kappa)|_B(B')=(p\otimes\kappa)(A\times B')=\int_A p(da)\cdot\kappa(a|B'),\hspace*{.5in}\forall B'\in {\cal B}(B).
\end{equation}

Suppose pre-action environment $\sigma_t\in {\cal P}(S)$ has been given for some period $t=1,...,\bar t$. Then, for every player with starting state $s_t$ in the period, his random action will be sampled from the distribution $\chi_t(s_t|\cdot)$ where as noted before, $\chi_t\in {\cal K}(S,X)$ is every player's behavioral guide. Thus, all players will together form the commonly felt in-action environment
\begin{equation}\label{consult1}
\tau_t=\sigma_t\otimes\chi_t.
\end{equation}
For each individual player with state $s_t$ and realized action $x_t$, his state $s_{t+1}$ in period $t+1$ will, by~(\ref{d-transition}), be distributed according to $\tilde g_t(s_t,x_t,\tau_t|\cdot)$. Thus, it will be reasonable for the pre-action environment in period $t+1$ to follow $\sigma_{t+1}=\tau_t\odot\tilde g_t(\cdot,\cdot,\tau_t)$, with
\begin{equation}\label{link}
[\tau_t\odot\tilde g_t(\cdot,\cdot,\tau_t)](S')=\int_{S\times X}\tau_t(ds\times dx)\cdot\tilde g_t(s,x,\tau_t|S'),\hspace*{.5in}\forall S'\in {\cal B}(S).
\end{equation}
Although~(\ref{link}) has been intuitively reasoned from~(\ref{d-transition}), we caution that logically it is part of the NG's definition rather than something derivable from the latter.

The transition from $\sigma_t$ to $\sigma_{t+1}$ through random action plan $\chi_t$ is best expressed by an operator. For any kernel $\chi\in {\cal K}(S,X)$, define operator $T_t(\chi)$ on the space ${\cal P}(S)$, so that
\begin{equation}\label{T-d}
T_t(\chi)\circ \sigma=(\sigma\otimes\chi)\odot\tilde g_t(\cdot,\cdot,\sigma\otimes\chi)=\sigma\odot\chi\odot\tilde g_t(\cdot,\cdot,\sigma\otimes\chi),\hspace*{.5in}\forall\sigma\in {\cal P}(S).
\end{equation}
Basically, state distribution $\sigma$ and random state-dependent action plan $\chi$ first fuse to form the joint state-action distribution $\sigma\otimes\chi$ to be felt by all players. The latter's random state transitions are then guided by the kernel $\tilde g_t(\cdot,\cdot,\sigma_t\otimes\chi)$. Subsequently, after ``averaging out'' impacts of actions, the next-period state distribution will become $\sigma\odot\chi\odot \tilde g_t(\cdot,\cdot,\sigma\otimes\chi)$. The one-period pre-action environment transition is now representable by
\begin{equation}\label{ng-onestep}
\sigma_{t+1}=T_t(\chi_t)\circ\sigma_t=\sigma_t\odot \chi_t\odot \tilde g_t(\cdot,\cdot,\sigma_t\otimes\chi_t).
\end{equation}

For periods $t$ and $t'$ with $t\leq t'$, as well as sequence $\chi_{[tt']}=(\chi_{t''}|t''=t,...,t')$ of action plans, we can iteratively define $T_{[tt']}(\chi_{[tt']})$, so that
\begin{equation}\label{okle}
T_{[tt']}(\chi_{[tt']})\circ \sigma_t=T_{t'}(\chi_{t'})\circ (T_{[t,t'-1]}(\chi_{[t,t'-1]})\circ \sigma_t),\hspace*{.5in}\forall \sigma_t\in {\cal P}(S).
\end{equation}
The left-hand side will be players' state distribution in period $t'+1$ when they start period $t$ with the distribution $\sigma_t$ and adopt the action sequence $\chi_{[tt']}$ in the interim. Note that $T_{[tt]}(\chi_{[tt]})$ is nothing but $T_t(\chi_t)$. As a default, we let $T_{[t,t-1]}$ stand for the identity operator on ${\cal P}(S)$.
The environment trajectory $\sigma_{[1,\bar t+1]}$ satisfies
\begin{equation}\label{sequence}
\sigma_{[1,\bar t+1]}=(T_{[1,t-1]}(\chi_{[1,t-1]})\circ\sigma_1\mid t=1,2,...,\bar t,\bar t+1).
\end{equation}
It is deterministic by definition.

\section{The $n$-player Game}\label{finite-game}

Let the same $\bar t$, $S$, $X$, $(\tilde f_t|t=1,2,...,\bar t)$, and $(\tilde g_t|t=1,2,...,\bar t)$ remain in the background. For some $n\in \mathbb{N}\setminus\{1\}$ and initial multi-state $s_1=(s_{11},s_{12},...,s_{1n})\in S^n$, we can define an $n$-player game $\Gamma_n(s_1)$, in which each $s_{1m}\in S$ is player $m$'s initial state. The game's payoffs and state evolutions are still described by the $\tilde f_t$'s and $\tilde g_t$'s, respectively. However, details are messier as outside environments vary from player to player and their evolutions are random. 

For $a\in A$, where $A$ is again a separable metric space, we use $\delta_a$ to denote the singleton Dirac measure with $\delta_a(\{a\})=1$. For $a=(a_1,...,a_n)\in A^n$ where $n\in \mathbb{N}$, we use $\varepsilon_a$ for $\sum_{m=1}^n \delta_{a_m}/n$, the empirical distribution generated by the vector $a$. We also use ${\cal P}_n(A)$ to denote the space of probability measures of the type $\varepsilon_a$ for $a\in A^n$, i.e., the space of empirical distributions generated from $n$ samples. Now back at the game $\Gamma_n(s_1)$, suppose in period $t=1,2,...,\bar t$, each player $m=1,2,...,n$ is in state $s_{tm}$ and takes action $x_{tm}$. Then, the in-action environment experienced by player 1 will be $\varepsilon_{s_{t,-1}x_{t,-1}}=\varepsilon_{((s_{t2},x_{t2}),...,(s_{tn},x_{tn}))}$. Thus, this player will receive payoff $\tilde f_t(s_{t1},x_{t1},\varepsilon_{s_{t,-1}x_{t,-1}})$ in the period, and his period-$(t+1)$ state $s_{t+1,1}$ will be sampled from the distribution $\tilde g_t(s_{t1},x_{t1},\varepsilon_{s_{t,-1}x_{t,-1}}|\cdot)$.

Suppose $\chi_{[1\bar t]}=(\chi_t\mid t=1,...,\bar t)\in ({\cal K}(S,X))^{\bar t}$ again describes the policy adopted by all $n$ players. 
Unlike in an NG, this time $\chi_{[1\bar t]}$ will help to generate a stochastic as opposed to deterministic environment trajectory. To describe each one-period transition in this complex process, we rely on the kernel $\chi_t^{\;n}\odot \tilde g_t^{\;n}\in {\cal K}(S^n,S^n)$ defined by
\begin{equation}\label{bh-def}
(\chi_t^{\;n}\odot \tilde g_t^{\;n})(s|S')=\int_{X^n}\chi_t^{\;n}(s|dx)\cdot \tilde g_t^{\;n}(s,x|S'),\hspace*{.5in}\forall s\in S^n, S'\in {\cal B}(S^n),
\end{equation}
where $\chi_t^{\;n}$ is a member of ${\cal K}(S^n,X^n)$ that satisfies
\begin{equation}\label{kn-def}
\chi_t^{\;n}(s|X'_1\times\cdots\times X'_n)=\Pi_{m=1}^n\chi_t(s_m|X'_m),\hspace*{.5in}\forall s\in S^n,X'_1,...,X'_n\in {\cal B}(X),
\end{equation}
and $\tilde g_t^{\;n}$ is a member of ${\cal K}(S^n\times X^n,S^n)$ that satisfies
\begin{equation}\label{gn-def}\begin{array}{l}
\tilde g_t^{\;n}(s,x|S'_1\times\cdots\times S'_n)=\Pi_{l=1}^n\tilde g_t(s_l,x_l,\varepsilon_{s_{-l}x_{-l}}|S'_l),\\
\hspace*{.5in}\forall (s,x)\in S^n\times X^n,S'_1,...,S'_n\in {\cal B}(S).
\end{array}\end{equation}
In combination,~(\ref{bh-def}) can be spelled out as
\begin{equation}
(\chi_t^{\;n}\odot \tilde g_t^{\;n})(s|S'_1\times\cdots\times S'_n)=\int_{X^n}\Pi_{m=1}^n \chi_t(s_m|dx_m)\cdot\Pi_{l=1}^n \tilde g_t(s_l,x_l,\varepsilon_{s_{-l}x_{-l}}|S'_l).
\end{equation}
The above reflects that, each player $m$ samples his action $x_m$ from the distribution $\chi_t(s_m|\cdot)$; once all players' actions $x=(x_1,...,x_n)$ have been determined, each player $l$ will face his unique in-action environment $\varepsilon_{s_{-l}x_{-l}}$; thus, this player's period-$(t+1)$ state will be sampled from the distribution $\tilde g_t(s_l,x_l,\varepsilon_{s_{-l}x_{-l}}|\cdot)$.

When the $n$ players start period $t$ with a random multi-state with distribution $\pi_{nt}\in {\cal P}(S^n)$ and they act according to random rule $\chi_t\in {\cal K}(S,X)$ in the period, they will generate the joint distribution $\mu_{nt}\in {\cal P}(S^n\times X^n)$ of period-$t$ multi-state and -action satisfying
\begin{equation}\label{consult2}
\mu_{nt}=\pi_{nt}\otimes\chi_t^{\;n}.
\end{equation}
According to~(\ref{s-to-mu}) and~(\ref{kn-def}), the above means that, for any $S'\in {\cal B}(S^n)$ and $X'_1,...,X'_n\in {\cal B}(X)$,
\begin{equation}\label{consult3}
\mu_{nt}(S'\times X'_1\times\cdots\times X'_n)=\int_{S'}\pi_{nt}(ds)\cdot \chi_t^{\;n}(s|X'_1\times\cdots\times X'_n)=\int_{S'}\pi_{nt}(ds)\cdot \Pi_{m=1}^n \chi_t(s_m|X'_m).
\end{equation}
Clearly,~(\ref{consult2}) corresponds to~(\ref{consult1}) in the NG situation.


By~(\ref{bh-def}), the period-$(t+1)$ multi-state distribution $\mu_{nt}\odot \tilde g_t^{\;n}\in {\cal P}(S^n)$ will follow
\begin{equation}\label{nnew-def}
(\mu_{nt}\odot \tilde g_t^{\;n})(S')=\int_{S^n\times X^n}\mu_{nt}(ds\times dx)\cdot \tilde g_t^{\;n}(s,x|S'),\hspace*{.5in}\forall S'\in {\cal B}(S^n).
\end{equation}
Combining~(\ref{consult2}) and~(\ref{nnew-def}), we can see that the one-period transition between multi-states is
\begin{equation}\label{finite-onestep}
\pi_{n,t+1}=(\pi_{nt}\otimes\chi_t^{\;n})\odot \tilde g_t^{\;n}=\pi_{nt}\odot \chi_t^{\;n}\odot\tilde g_t^{\;n}.
\end{equation}
Note~(\ref{finite-onestep}) is the $n$-player game's answer to the NG's~(\ref{ng-onestep}). Similar to~(\ref{okle}), for $t\leq t'$, the distribution $\pi_{nt'}$ of period-$t'$ multi-state $s_{t'}$ is given by
\begin{equation}\label{muster}
\pi_{nt'}=\pi_{nt}\odot\Pi_{t''=t}^{t'-1}(\chi_{t''}^{\;n}\odot\tilde g_{t''}^{\;n}).
\end{equation}
When the initial multi-state $s_1$ is randomly drawn from distribution $\pi_{n1}$, the entire trajectory $\pi_{n,[1,\bar t+1]}=(\pi_{nt}|t=1,2,...,\bar t,\bar t+1)$ of the $n$-player game's multi-state distributions can be written as
\begin{equation}\label{sequence-n}
\pi_{n,[1,\bar t+1]}=(\pi_{n1}\odot\Pi_{t'=1}^{t-1}(\chi_{t'}^{\;n}\odot\tilde g_{t'}^{\;n})|t=1,2,..,\bar t,\bar t+1).
\end{equation}
When all players' states are sampled from some $\sigma_1\in {\cal P}(S)$, we still have~(\ref{sequence-n}) as the trajectory for multi-state distributions, but with $\pi_{n1}=\sigma_1^{\;n}$.
When recognizing $\pi_{n1}=\delta_{s_1}$, the Dirac measure in ${\cal P}(S^n)$ that assigns the full weight to $s_1$,~(\ref{sequence-n}) will help describe the evolution of the multi-state distribution for the $n$-player game $\Gamma_n(s_1)$, much like~(\ref{sequence}) did for $\Gamma(\sigma_1)$.

\section{Convergence of Aggregate Environments}\label{convergence}

Even before touching upon notions like cumulative payoffs and equilibria, we can already introduce an interesting link between finite games and NGs. It is in terms of an asymptotic relationship between a sequence $\pi_{n,[t,\bar t+1]}=(\pi_{nt'}|t'=t,t+1,...,\bar t+1)$ of multi-state distributions in $n$-player games and a sequence $\sigma_{[t,\bar t+1]}=(\sigma_{t'}|t'=t,t+1,...,\bar t+1)$ of state distributions in their NG counterparts. The message is that, when starting from  similar environments in period $t$ and adopting the same action plan from that period on, stochastic environment paths experienced by large finite games will not drift too much away from the NG's deterministic environment trajectory. We refrain from using the word convergence because the $\pi_{nt'}$'s reside in different spaces for different $n$'s.

First, we propose the concept asymptotic resemblance in order to precisely describe the way in which members in a sequence of probability measures increasingly resemble the products of a given measure. 
For a separable metric space $A$, the space ${\cal P}(A)$ is metrized by the Prohorov metric $\rho_A$, which induces the weak topology on it. At fixed $n\in \mathbb{N}$, the map $\varepsilon_{(\cdot)}$ from $A^n$ to ${\cal P}_n(A)\subseteq {\cal P}(A)$ is continuous. Therefore, for any $p\in {\cal P}(A)$ and $\epsilon>0$, the set $\{a\in A^n|\rho_A(\varepsilon_a,p)<\epsilon\}$ is an open subset of $A^n$ and thus a member of ${\cal B}(A^n)$.

{\definition\label{conv-p} For a separable metric space $A$, suppose $p\in {\cal P}(A)$ and for each $n\in \mathbb{N}$, $q_n\in {\cal P}(A^n)$. We say that sequence $q_n$ asymptotically resembles the sequence $p^n$ made up of $p$'s $n$-th order products $p\times\cdots\times p$, if for any $\epsilon>0$ and $n$ that is large enough,
\[ q_n(\{a\in A^n|\rho_A(\varepsilon_a,p)<\epsilon\})>1-\epsilon. \]}Definition~\ref{conv-p} says that sequence $q_n$ will asymptotically resemble the sequence $p^n$ of product measures when the empirical distribution $\varepsilon_a$ of a random vector $a=(a_1,...,a_n)$, sampled from $q_n$, is highly likely to be close to $p$ as $n$ approaches $+\infty$. This resemblance notion is consistent with Prohorov's theorem (Parthasarathy \cite{P05}, Theorem II.7.1), whose weak version is presented as Lemma~\ref{p-Prohorov} in Appendix~\ref{app-a}. Due to it, any sequence $(p')^n$ will asymptotically resemble the sequence $p^n$ if and only if $p'=p$.

Some results related to the resemblance concept have been placed in Appendix~\ref{app-a}. Lemma~\ref{p-uniform} stems from Dvoretzky, Kiefer, and Wolfolwitz's \cite{DKW56} inequality and makes the convergence in Lemma~\ref{p-Prohorov} uniform in the chosen probability $p$. According to Lemma~\ref{p-mc8}, the tampering of one component within any $n$-long vector $a\in A^n$ would not much alter $\varepsilon_a$. It is therefore natural for Lemma~\ref{d-prob} to state that the resemblance of $q_n$ to $p^n$ would lead to that of the $A^{n-1}$-marginal $q_n|_{A^{n-1}}$ to $p^{n-1}$. Lemma~\ref{dc-prob} says that the above would also lead to the asymptotic resemblance of $p'\times q_{n-1}$ to $p^n$ for any $p'$. So in general there can be nothing substantial regarding the relationship between the $A$-marginals $q_n|_A$ and $p$. 
Finally, Lemma~\ref{c-prob} shows that asymptotic resemblance is preserved under the projection of $A\times B$ into $A$.

The following one-step result states that asymptotic resemblance concerning pre-action environments is translatable into that concerning in-action environments; also, the same resemblance is preserved after undergoing one single step in a game.

\begin{proposition}\label{T-onestep}
Let state distribution $\sigma\in {\cal P}(S)$, random state-dependent action plan $\chi\in {\cal K}(S,X)$, and state-transition kernel $g\in {\cal G}(S,X)$, with the latter enjoying the continuity of $g(s,x,\tau)$ in the joint state-action distribution $\tau$ at an $(s,x)$-independent rate. Also, multi-state distribution $\pi_n\in {\cal P}(S^n)$ for each $n\in \mathbb{N}$. Suppose further that the sequence $\pi_n$ asymptotically resembles the sequence $\sigma^n$. Then, \\
\indent\M (i) the sequence $\pi_n\otimes\chi^n$ will asymptotically resemble the sequence $(\sigma\otimes\chi)^n$, and \\
\indent\M (ii) the sequence $\pi_n\odot \chi^n\odot g^n$ will asymptotically resemble the sequence $(\sigma\odot \chi\odot g(\cdot,\cdot,\sigma\otimes\chi))^n$.\\
Indeed, (ii) remains valid under mild contamination. That is, for any $(s,x)\in S\times X$,\\
\indent\M (iii) the sequence $(\delta_{sx}\times (\pi_{n-1}\otimes \chi^{n-1}))\odot g^n$ will asymptotically resemble the sequence $(\sigma\odot \chi\odot g(\cdot,\cdot,\sigma\otimes\chi))^n$ at a rate independent of the chosen $(s,x)$.
\end{proposition}

Proposition~\ref{T-onestep} is one of our two most technical results. Its proof invokes both Prohorov's theorem (Parthasarathy \cite{P05}, Theorem II.7.1) on the convergence of empirical distributions and for parts (ii) and (iii), Dvoretzky, Kiefer, and Wolfolwitz's \cite{DKW56} inequality which provides the uniformity of such convergence. In the proposition, part (i) stresses the passibility from convergence of pre-action environments to that of same-period in-action environments, see~(\ref{consult1}) and~(\ref{consult2}); part (ii) further points out that convergence in next-period pre-action environments will follow suit, see~(\ref{ng-onestep}) and~(\ref{finite-onestep}); also, part (iii) will be useful when we take the view point from one single player..

To take advantage of Proposition~\ref{T-onestep}, we now assume the equi-continuity of the state transitions with respect to in-action environments.


{\assumption\label{g-c} Each transition kernel $\tilde g_t(s,x,\tau)$ is continuous in $\tau$ at an $(s,x)$-independent rate. That is, for any in-action environment $\tau\in {\cal P}(S\times X)$ and $\epsilon>0$, there is $\delta>0$, such that for any $\tau'\in {\cal P}(S\times X)$ satisfying $\rho_{S\times X}(\tau,\tau')<\delta$ and any $(s,x)\in S\times X$,
	\[ \rho_S(\tilde g_t(s,x,\tau),\tilde g_t(s,x,\tau'))<\epsilon. \]}We are in a position to derive this section's main result. It states that, when an NG and its finite counterparts evolve under the same action plan, environment pathways of large finite games, though stochastic, will resemble the deterministic pathway of the NG.

\begin{theorem}\label{T-converge}
Let a policy profile $\chi_{[t\bar t]}\in ({\cal K}(S,X))^{\bar t-t+1}$ for periods $t,t+1,...,\bar t$ be given. When $s_t=(s_{t1},...,s_{tn})$ has a distribution $\pi_{nt}$ that asymptotically resembles $\sigma_t^{\;n}$, the series $(\pi_{nt}\odot\Pi_{t''=t}^{t'-1}(\chi_{t''}^{\;n}\odot\tilde g_{t''}^{\;n})\mid t'=t,t+1,...,\bar t,\bar t+1)$ will asymptotically resemble $((T_{[t,t'-1]}(\chi_{[t,t'-1]})\circ\sigma_t)^n\mid t'=t,t+1,...,\bar t,\bar t+1)$ as well. That is, for any $\epsilon>0$ and any $n$ that is large enough,
\[ [\pi_{nt}\odot \Pi_{t''=t}^{t'-1}(\chi_{t''}^{\;n}\odot\tilde g_{t''}^{\;n})](\tilde A_{nt'}(\epsilon))>1-\epsilon,\hspace*{.5in}\forall t'=t,t+1,...,\bar t+1,\]
where for each $t'$, the set of multi-states $\tilde A_{nt'}(\epsilon)\in {\cal B}(S^n)$ is such that,
\[ \rho_S(\varepsilon_{s_{t'}},T_{[t,t'-1]}(\chi_{[t,t'-1]})\circ\sigma_t)<\epsilon, \hspace*{.5in}\forall s_{t'}\in \tilde A_{nt'}(\epsilon). \]
\end{theorem}

Suppose an NG starts period $t$ with pre-action environment $\sigma_t$ and a slew of finite games start the period with pre-action environments that are ever nearly sampled from $\sigma_t$. Let the evolution of both types of games be guided by players acting according to the same policy profile $\chi_{[t\bar t]}$. Then, as the numbers of players $n$ involved in finite games grow indefinitely, Theorem~\ref{T-converge} predicts for ever less chances for the finite games' period-$t'$ environments $\varepsilon_{s_{t'}}$ to be even slightly away from the NG's deterministic period-$t'$ environment $T_{[t,t'-1]}(\chi_{[t,t'-1]})\circ\sigma_t$. For some fixed $\sigma_1\in {\cal P}(S)$, we can plug $t=1$ and $\pi_{n1}=\sigma_1^{\;n}$ into Theorem~\ref{T-converge}. Then, we will obtain the proximity between $\sigma^{\;n}_{[1,\bar t+1]}=(\sigma^{\;n}_t|t=1,2,...,\bar t,\bar t+1)$ and $\pi_{n,[1,\bar t+1]}=(\pi_{nt}|t=1,2,...,\bar t,\bar t+1)$ for large $n$'s, where every $\sigma_t=T_{[1,t-1]}(\chi_{[1,t-1]})\circ\sigma_1$ and every $\pi_{nt}=\sigma_1^{\;n}\odot\Pi_{t'=1}^{t-1}(\chi_{t'}^{\;n}\odot\tilde g_{t'}^{\;n})$. In view of~(\ref{sequence}) and~(\ref{sequence-n}), this means that when large games sample their initial states from an NG's starting distribution $\sigma_1$, the former games' state-distribution trajectories will remain close to that of the latter game.

Our confinement so far to discrete spaces $S$ and $X$ arises mainly from the need to deal with non-product joint probabilities of the form $p\otimes \kappa$; see~(\ref{s-to-mu}). In Yang \cite{Y12}, where random state transitions and random action plans were modeled through independently generated shocks, only results pertaining to product-form probabilities $p\times q$, where $q$ is an ordinary rather than conditional probability, were needed. Because of this,
known properties like Propositions III.4.4 and III.4.6 of Ethier and Kurtz \cite{EK86} could be put to good use. Results there could thus be based on complete state and shock spaces. In contrast, if we were to consider more general spaces here, we would face the presently unsurmountable challenge of passing the closeness between measures $p$ and $p_i$ for $i=1,2,...,n$ onto that between $p^n$ and $\prod_{i=1}^n p_i$ when $n$ itself tends to infinity.

\section{NG and Finite-game Equilibria}\label{equilibria}

We present this paper's main result that an NG equilibrium, though oblivious of past history and blind to other players' states, will generate minimal regrets when adopted by players in large finite games. First, we introduce equilibrium concepts used in both types of games.

\subsection{Equilibria in NG}

In defining the NG $\Gamma(\sigma_1)$'s equilibria, we subject a candidate policy profile to one-time
deviation of a single player, who is by default infinitesimal in influence.
Note the deviation will not alter the environment trajectory corresponding to the candidate profile. With this understanding, we define $v_t(s_t,\xi_{[t\bar t]},\sigma_t,\chi_{[t\bar t]})$ as the total expected payoff a player can receive from period $t$ to $\bar t$, when he starts with state $s_t\in S$ and adopts action plan $\xi_{[t\bar t]}=(\xi_t,...,\xi_{\bar t})\in ({\cal K}(S,X))^{\bar t-t+1}$ throughout, while other players form initial pre-action environment $\sigma_t\in {\cal P}(S)$ and adopt policy profile $\chi_{[t\bar t]}=(\chi_t,...,\chi_{\bar t})\in ({\cal K}(S,X))^{\bar t-t+1}$ throughout. As a terminal condition, we certainly have
\begin{equation}\label{terminal}
v_{\bar t+1}(s_{\bar t+1},\sigma_{\bar t+1})=0.
\end{equation}
For $t=\bar t,\bar t-1,...,1$, we have the recursive relationship
\begin{equation}\label{recursive}\begin{array}{ll}
v_t(s_t,\xi_{[t\bar t]},\sigma_t,\chi_{[t\bar t]})=\int_X \xi_t(s_t|dx_t)\cdot[\tilde f_t(s_t,x_t,\sigma_t\otimes \chi_t)\\
\;\;\;\;\;\;\;\;\;\;\;\;\;\;\;\;\;\;+\int_S \tilde g_t(s_t,x_t,\sigma_t\otimes \chi_t|ds_{t+1})\cdot v_{t+1}(s_{t+1},\xi_{[t+1,\bar t]},T_t(\chi_t)\circ \sigma_t,\chi_{[t+1,\bar t]})].
\end{array}\end{equation}
This is because the player's action is guided in a random fashion by $\xi_t$, its payoff is determined by $\tilde f_t$, its state evolution is governed by $\tilde g_t$, and its future payoff is supplied by $v_{t+1}$; also, after undergoing the commonly adopted action plan $\chi_t$, the period-$(t+1)$ pre-action environment $\sigma_{t+1}$ will be $T_t(\chi_t)\circ\sigma_t$ as shown in~(\ref{ng-onestep}). The choice of $\xi_t$ affects the current player's period-$t$ action $x_t$, his period-$(t+1)$ state $s_{t+1}$, and his future state-action trajectory. However, the change at this negligible player does not alter the period-$t$ in-action environment $\sigma_t\otimes\chi_t$ as listed in~(\ref{consult1}) or any environment in the future. This is the main reason why NGs are easier to handle than their finite-player counterparts.

Now, we deem policy $\chi_{[1\bar t]}\in ({\cal K}(S,X))^{\bar t}$ a Markov equilibrium for the game $\Gamma(\sigma_1)$ when, for every $t=1,2,...,\bar t$ and $\xi_t\in {\cal K}(S,X)$,
\begin{equation}\label{first}
v_t(s_t,\chi_{[t\bar t]},\sigma_t,\chi_{[t\bar t]})\geq v_t(s_t,(\xi_t,\chi_{[t+1,\bar t]}),\sigma_t,\chi_{[t\bar t]}),\hspace*{.5in}\forall s_t\in S,
\end{equation}
where
\begin{equation}\label{environ}
\sigma_t=T_{[1,t-1]}(\chi_{[1,t-1]})\circ \sigma_1.
\end{equation}
That is, policy $\chi_{[1\bar t]}$ will be regarded an equilibrium when no player can be better off by unilaterally deviating to any alternative plan $\xi_t\in {\cal K}(S,X)$ in any single period $t$. The definition of $\sigma_t$ in~(\ref{environ}) underscores the evolution of the deterministic environment trajectory following the adoption of action plan $\chi_{[1\bar t]}$ by almost all players. 

\subsection{$\epsilon$-Equilibria in $n$-player Games}

For an $n$-player game, let $v_{nt}(s_{t1},\xi_{[t\bar t]},\varepsilon_{s_{t,-1}},\chi_{[t\bar t]})$ be the total expected payoff player 1 can receive from period $t$ to $\bar t$, when he starts with state $s_{t1}\in S$ and adopts action plan $\xi_{[t\bar t]}\in ({\cal K}(S,X))^{\bar t-t+1}$ throughout, while other players form initial empirical state distribution $\varepsilon_{s_{t,-1}}=\varepsilon_{(s_{t2},...,s_{tn})}\in {\cal P}_{n-1}(S)$ and adopt action plan $\chi_{[t\bar t]}\in ({\cal K}(S,X))^{\bar t-t+1}$ throughout. As a terminal condition, we have
\begin{equation}\label{terminal-n}
v_{n,\bar t+1}(s_{\bar t+1,1},\varepsilon_{s_{\bar t+1,-1}})=0.
\end{equation}
For $t=\bar t,\bar t-1,...,1$, we have the recursive relationship
\begin{equation}\label{recursive-n}\begin{array}{l}
v_{nt}(s_{t1},\xi_{[t\bar t]},\varepsilon_{s_{t,-1}},\chi_{[t\bar t]})=\int_X\xi_t(s_{t1}|dx_{t1})\cdot \int_{X^{n-1}}\chi_t^{\;n-1}(s_{t,-1}|dx_{t,-1})\times\\
\;\;\times[\tilde f_t(s_{t1},x_{t1},\varepsilon_{s_{t,-1}x_{t,-1}})
+\int_{S^n}\tilde g_t^{\;n}(s_t,x_t|ds_{t+1})\cdot v_{n,t+1}(s_{t+1,1},\xi_{[t+1,\bar t]},\varepsilon_{s_{t+1},-1},\chi_{[t+1,\bar t]})],
\end{array}\end{equation}
where the meaning of $\chi_t^{\;n-1}(s_{t,-1}|dx_{t,-1})$ follows from~(\ref{kn-def}) and that of $\tilde g_t^{\;n}(s_t,x_t|ds_{t+1})$ follows from~(\ref{gn-def}). Note~(\ref{recursive-n}) differs substantially from its NG counterpart~(\ref{recursive}). With only a finite number of players, player 1's one-time choice $\xi_t$ not only affects his own future actions and states as before, but differently, starting from the altered in-action environment $\varepsilon_{s_tx_t}$, it also impacts the entire future trajectory of all other players. Note $\varepsilon_{s_tx_t}$ impacts the generation of $s_{t+1}=(s_{t+1,1},...,s_{t+1,n})$ in its projections to $n$ different $(n-1)$-dimensional spaces, as according to~(\ref{gn-def}), $\int_{S^n}\tilde g_t^{\;n}(s_t,x_t|ds_{t+1})$ amounts to $\Pi_{m=1}^n\int_S \tilde g_t(s_{tm},x_{tm},\varepsilon_{s_{t,-m}x_{t,-m}}|ds_{t+1,m})$.

For each $n\in \mathbb{N}\setminus\{1\}$, let $\hat\pi_{n-1,[1\bar t]}=(\hat\pi_{n-1,t}\mid t=1,...,\bar t)\in ({\cal P}(S^{n-1}))^{\bar t}$ be a series of other-player multi-state distributions. For $\epsilon\geq 0$, we deem $\chi_{[1\bar t]}=(\chi_t\mid t=1,...,\bar t)\in ({\cal K}(S,X))^{\bar t}$ an $\epsilon$-Markov equilibrium for the game family $(\Gamma_n(s_1)\mid s_1\in S^n)$ in the sense of $\hat\pi_{n-1,[1\bar t]}$ when, for every $t=1,...,\bar t$, $\xi_{[t\bar t]}\in ({\cal K}(S,X))^{\bar t-t+1}$, and $s_{t1}\in S$,
\begin{equation}\label{second}\begin{array}{l}
 \int_{S^{n-1}}\hat\pi_{n-1,t}(ds_{t,-1})\cdot v_{nt}(s_{t1},\chi_{[t\bar t]},\varepsilon_{s_{t,-1}},\chi_{[t\bar t]})\\
  \;\;\;\;\;\;\;\;\;\;\;\;\;\;\;\;\;\;\;\;\;\;\;\;\geq \int_{S^{n-1}}\hat\pi_{n-1,t}(ds_{t,-1})\cdot v_{nt}(s_{t1},\xi_{[t\bar t]},\varepsilon_{s_{t,-1}},\chi_{[t\bar t]})-\epsilon.
\end{array}\end{equation}
That is, action plan $\chi_{[1\bar t]}$ will be an $\epsilon$-Markov equilibrium in the sense of $\hat\pi_{n-1,[1\bar t]}$ when under the plan's guidance, the average payoff from any period $t$ and player-1 state $s_{t1}$ on cannot be improved by more than $\epsilon$ through any unilateral deviation, where the ``average'' is based on other players' multi-state $s_{t,-1}$ being sampled from the distribution $\hat\pi_{n-1,t}$. Note~(\ref{second}) differs from~(\ref{first}) also in that its unilateral deviation need not be one-time.

\subsection{Main Transient Result}

Before moving on, we need the single-period payoff functions $\tilde f_t$ to be continuous.

{\assumption\label{f-c}
Each payoff function $\tilde f_t(s,x,\tau)$ is continuous in the in-action environment $\tau$ at an $(s,x)$-independent rate. That is, for any $\tau\in {\cal P}(S\times X)$ and $\epsilon>0$, there is $\delta>0$, such that for any $\tau'\in {\cal P}(S\times X)$ satisfying $\rho_{S\times X}(\tau,\tau')<\delta$ and any $(s,x)\in S\times X$,
\[ \mid \tilde f_t(s,x,\tau)-\tilde f_t(s,x,\tau')\mid<\epsilon. \]}Now we show the convergence of finite-game value functions to their NG counterpart, the proof of which is quite technical as well, and calls upon parts (i) and (iii) of Proposition~\ref{T-onestep}.

\begin{proposition}\label{V-convergence}
For any $t=1,2,...,\bar t+1$, let $\sigma_t\in {\cal P}(S)$ and $\hat\pi_{n-1,t}\in {\cal P}(S^{n-1})$ for each $n\in \mathbb{N}$. Suppose the sequence $\hat\pi_{n-1,t}$ asymptotically resembles the sequence $\sigma_t^{\;n-1}$. Then for any $\chi_{[t\bar t]}\in ({\cal K}(S,X))^{\bar t-t+1}$, the sequence $\int_{S^{n-1}}\hat\pi_{n-1,t}(ds_{t,-1})\cdot v_{nt}(s_{t1},\xi_{[t\bar t]},\varepsilon_{s_{t,-1}},\chi_{[t\bar t]})$ will converge to $v_t(s_{t1},\xi_{[t\bar t]},\sigma_t,\chi_{[t\bar t]})$ at a rate that is independent of both $s_{t1}\in S$ and $\xi_{[t\bar t]}\in ({\cal K}(S,X))^{\bar t-t+1}$.
\end{proposition}

Combining~(\ref{first}) and~(\ref{second}), as well as Proposition~\ref{V-convergence}, we can come to the main result.

\begin{theorem}\label{main}
For some $\sigma_1\in {\cal P}(S)$, suppose $\chi_{[1\bar t]}=(\chi_t\mid t=1,2,...,\bar t)\in ({\cal K}(S,X))^{\bar t}$ is a Markov equilibrium of NG $\Gamma(\sigma_1)$. Also, suppose $\hat\pi_{n-1,[1\bar t]}=(\hat\pi_{n-1,t}|t=1,2,...,\bar t)\in ({\cal P}(S^{n-1}))^{\bar t}$ is such that the sequence $\hat\pi_{n-1,t}$ asymptotically resembles the sequence $\sigma_t^{\;n-1}$ for each $t$, where $\sigma_t=T_{[1,t-1]}(\chi_{[1,t-1]})\circ \sigma_1$. Then, for $\epsilon>0$ and large enough $n\in \mathbb{N}$, the given $\chi_{[1\bar t]}$ is also an $\epsilon$-Markov equilibrium for the game family $(\Gamma_n(s_1)\mid s_1\in S^n)$ in the sense of $\hat\pi_{n-1,[1\bar t]}$.
\end{theorem}


The theorem says that players in a large finite game can agree on an NG equilibrium and expect to lose little on average, as long as the other-player multi-state distribution $\hat\pi_{n-1,t}$ on which ``average'' is based is similar to the product form $\sigma_t^{\;n-1}$, where $\sigma_t=T_{[1,t-1]}(\chi_{[1,t-1]})\circ \sigma_1$ is the corresponding NG's predictable equilibrium state distribution for the same period. As to whether reasonable $\hat\pi_{n-1,[1\bar t]}=(\hat\pi_{n-1,t}|t=1,2,...,\bar t)$ exists to satisfy this condition, the answer is affirmative. The next section is dedicated to this point.

\section{The Condition in Theorem~\ref{main}}\label{interpretation}

We now present examples where the key condition in Theorem~\ref{main} can be true. 
In all of them, we let the initial other-player multi-state distribution $\hat\pi_{n-1,1}=\sigma_1^{\;n-1}=\sigma_1^{\;n}|_{S^{n-1}}$. That is, we let players' initial states in $n$-player games be randomly drawn from the NG's initial state distribution $\sigma_1$.
Now we discuss what can happen in periods $t=2,3,...,\bar t$.

\subsection{Two Possibilities}

First, we can let each $\hat\pi_{n-1,t}=\sigma_t^{\;n-1}$. It has been discussed right after Definition~\ref{conv-p} that the sequence $\sigma_t^{\;n-1}$ asymptotically resembles itself. So this choice satisfies the condition in Theorem~\ref{main}. This would correspond to the case where players in large finite games take the ``lazy'' approach of using independent draws on the NG state distribution to assess their opponents' states. Note this is reasonable due to the common initial condition for both types of games and Theorem~\ref{T-converge}.

Second, we can let each $\hat\pi_{n-1,t}=\pi_{nt}|_{S^{n-1}}$, where
\begin{equation}\label{second-e}
\pi_{nt}=\sigma_1^{\;n}\odot\Pi_{t'=1}^{t-1}(\chi_{t'}^{\;n}\odot\tilde g_{t'}^{\;n}).
\end{equation}
According to~(\ref{muster}), $\pi_{nt}$ stands for players' multi-state distribution in period $t$ in an $n$-player game when their initial states are randomly drawn from the distribution $\sigma_1$ and then from period 1 onward players all follow through with the NG equilibrium $\chi_{[1\bar t]}$. Since the sequence $\sigma_1^{\;n}$ asymptotically resembles itself, Theorem~\ref{T-converge} will ascertain the asymptotic resemblance of $\pi_{nt}$ to $\sigma_t^{\;n}$. Then, Lemma~\ref{d-prob} in Appendix~\ref{app-a} will lead to the asymptotic resemblance of $\hat\pi_{n-1,t}$ to $\sigma_t^{\;n-1}$. So this choice would satisfy Theorem~\ref{main}'s condition as well. Also, its meaning is clear---here players in large finite games use precise assessments on what other players' states might be had they followed the NG equilibrium all along.

\subsection{Refinement and a Third Choice}

Note that $\hat\pi_{n-1,t}$ has not countenanced the possibility in which a player involves his own state $s_{t1}$ in the estimation of the other-player multi-state $s_{t,-1}$. We now show that this is possible at least when the state space $S$ is finite. In that case, we can upgrade the $\hat\pi_{n-1,t}\in {\cal P}(S^{n-1})$ in Proposition~\ref{V-convergence} to $\hat\pi_{n-1,t}(\cdot)=(\hat\pi_{n-1,t}(s_{t1}|\cdot)|s_{t1}\in S)\in ({\cal P}(S^{n-1}))^S$ and obtain the convergence of $\int_{S^{n-1}}\hat\pi_{n-1,t}(s_{t1}|ds_{t,-1})\cdot v_{nt}(s_{t1},\xi_{[t\bar t]},\varepsilon_{s_{t,-1}},\chi_{[t\bar t]})$ to $v_t(s_{t1},\xi_{[t\bar t]},\sigma_t,\chi_{[t\bar t]})$ at an $s_{t1}$-independent rate. This will lead us to the following extended version of Theorem~\ref{main}. 

\begin{theorem}\label{main-f}
Suppose $\sigma_{[1,\bar t+1]}$ and $\chi_{[1\bar t]}$ are all the same as in Theorem~\ref{main}. Also, suppose $\hat\pi_{n-1,[1\bar t]}(\cdot)=(\hat\pi_{n-1,t}(s_{t1}|\cdot)|t=1,2,...,\bar t,\;s_{t1}\in S)\in (({\cal P}(S^{n-1}))^S)^{\bar t}$ is such that the sequence $\hat\pi_{n-1,t}(s_{t1}|\cdot)$ asymptotically resembles the sequence $\sigma_t^{\;n-1}$ for each $t$ and $s_{t1}$. Then, for $\epsilon>0$ and large enough $n\in \mathbb{N}$, for every $t=1,...,\bar t$, $\xi_{[t\bar t]}\in ({\cal K}(S,X))^{\bar t-t+1}$, and $s_{t1}\in S$,
\[\begin{array}{l}
  \int_{S^{n-1}}\hat\pi_{n-1,t}(s_{t1}|ds_{t,-1})\cdot v_{nt}(s_{t1},\chi_{[t\bar t]},\varepsilon_{s_{t,-1}},\chi_{[t\bar t]})\\
  \;\;\;\;\;\;\;\;\;\;\;\;\;\;\;\;\;\;\;\;\;\;\;\;\geq \int_{S^{n-1}}\hat\pi_{n-1,t}(s_{t1}|ds_{t,-1})\cdot v_{nt}(s_{t1},\xi_{[t\bar t]},\varepsilon_{s_{t,-1}},\chi_{[t\bar t]})-\epsilon.
  \end{array}\]
\end{theorem}

For it to satisfy the condition in Theorem~\ref{main-f}, we can still let $\hat\pi_{n-1,[1\bar t]}(\cdot)$ be the same as in the aforementioned two examples, in which the newly added $s_{t1}$-dependence is mute. But a third choice would allow each player a full-fledged Bayesian update on other players' states.

In this third choice, we still use~(\ref{second-e}) to define $\pi_{nt}$. Then, as long as $\sigma_t(s_{t1})>0$, we let
\begin{equation}\label{final-def}
\hat\pi_{n-1,t}(s_{t1}|\cdot)=\pi_{nt,S}|_{S^{n-1}}(s_{t1}|\cdot),
\end{equation}
the other-player multi-state distribution derivable from $\pi_{nt}$ when conditioned on the current player's state $s_{t1}$; otherwise, we simply let $\hat\pi_{n-1,t}=\pi_{nt}|_{S^{n-1}}$ just as in the second example. Note the marginal $\pi_{nt}|_S$ is defined by
\begin{equation}\label{marginal-pi}
\pi_{nt}|_S(\{s_{t1}\})=\pi_{nt}(\{s_{t1}\}\times S^{n-1}),\hspace*{.5in}\forall s_{t1}\in S,
\end{equation}
and each conditional distribution $\pi_{nt,S}|_{S^{n-1}}(s_{t1}|\cdot)$ is defined by
\begin{equation}\label{conditional-pi}
\pi_{nt,S}|_{S^{n-1}}(s_{t1}|S')=\frac{\pi_{nt}(\{s_{t1}\}\times S')}{\pi_{nt}|_S(\{s_{t1}\})}=\frac{\pi_{nt}(\{s_{t1}\}\times S')}{\pi_{nt}(\{s_{t1}\}\times S^{n-1})},\hspace*{.5in}\forall S'\in {\cal B}(S^{n-1}),
\end{equation}
when the denominator is strictly positive and an arbitrary value otherwise.

\subsection{Symmetry Makes it Work}

The lone fact that $\pi_{nt}$ asymptotically resembles $\sigma_t^{\;n}$ is actually quite far from being able to dictate the asymptotic resemblance of the thus defined $\hat\pi_{n-1,t}(s_{t1}|\cdot)$ to $\sigma_t^{\;n-1}$. Note that for a general $q_n$ resembling some $p^n$, Lemma~\ref{dc-prob} in Appendix~\ref{app-a} has all but ruled out the convergence of $\pi_n|_A$ to $p$, let alone the asymptotic resemblance of $q_{n,A}|_{A^{n-1}}$ to $p^{n-1}$. Fortunately, $\pi_{nt}$ still enjoys the additional feature of being symmetric.

For any $n\in\mathbb{N}$, let $\Psi_n$ be the set of all $n$-dimensional permutations. That is, each $\psi\in\Psi_n$ makes $(\psi(1),...,\psi(n))$ a permutation of $(1,...,n)$. For a given $\psi\in\Psi_n$, let us suppose $\psi a=(a_{\psi(1)},...,a_{\psi(n)})$ for any $a=(a_1,...,a_n)\in A^n$, and then $\psi A'=\{\psi a|a\in A'\}$ for any $A'\subseteq A^n$. Note that, due to its innately symmetric definition, ${\cal B}(A^n)$ is automatically symmetric in the sense that ${\cal B}(A^n)=\{\psi A'|A'\in {\cal B}(A^n)\}$ for any $\psi\in\Psi_n$.

{\definition\label{sym-def} For $n\in \mathbb{N}$ and separable metric space $A$, we say $q_n\in {\cal P}(A^n)$ symmetric if
\[ q_n(A')=q_n(\psi A'),\hspace*{.5in}\forall \psi\in \Psi_n,\;A'\in {\cal B}(A^n).\]}

We have the much needed result that asymptotic resemblance of $q_n$ to $p^n$ does lead to the convergence of $q_n|_A$ to $p$ when $q_n$ is symmetric. This is in stark contrast with Lemma~\ref{dc-prob}.

\begin{proposition}\label{bomb1}
Let $A$ be a discrete metric space and $q_n\in {\cal P}(A^n)$ for every $n\in \mathbb{N}$ be symmetric. Suppose the sequence $q_n$ asymptotically resembles the sequence $p^n$. Then, the sequence $q_n|_A$ will converge to $p$, namely, $\lim_{n\rightarrow +\infty}q_n|_A(\{a\})=p(\{a\})$ for every $a\in A$.
\end{proposition}

This then results in the resemblance of $q_{n,A}|_{A^{n-1}}$ to $p^{n-1}$.

\begin{proposition}\label{bomb2}
Let $A$ be a discrete metric space and $q_n\in {\cal P}(A^n)$ for every $n\in \mathbb{N}$ be symmetric. Suppose the sequence $q_n$ asymptotically resembles the sequence $p^n$. Then, the sequence $q_{n,A}|_{A^{n-1}}(a|\cdot)$ will asymptotically resemble the sequence $p^{n-1}$ for any $a\in A$ with $p(\{a\})>0$. 
\end{proposition}

Note that $\pi_{n1}$, being equal to $\sigma_1^{\;n}$, is symmetric. As suggested by~(\ref{second-e}), the operation it has to go through to arrive to $\pi_{nt}$ is also symmetric. Hence, $\pi_{nt}$ is symmetric. Therefore, by Proposition~\ref{bomb1}, the marginal probability $\pi_{nt}|_S$ as defined in~(\ref{marginal-pi}) would converge to the NG state distribution $\sigma_t$; thus, the conditional distribution $\pi_{nt,S}|_{S^{n-1}}(s_{t1}|\cdot)$ as defined in~(\ref{conditional-pi}) would be well defined when $\sigma_t(s_{t1})>0$. Then, Proposition~\ref{bomb2} can guarantee that $\hat\pi_{n-1,t}(s_{t1}|\cdot)$ as defined in~(\ref{final-def}) would asymptotically resemble $\sigma_t^{\;n-1}$ and hence help to facilitate the condition needed for Theorem~\ref{main-f}. The above suggests that, even when players exercise the most accurate Bayesian updates on other players' states using their own state information, they will not discern much regret on average by adhering to the NG equilibrium. 

\section{A Stationary Situation}\label{stationary}


Now we study an infinite-horizon model with stationary features. To this end, we keep $S$ and $X$, but let there be a discount factor $\bar\alpha\in [0,1)$. There is a payoff function $\tilde f$ which meets the basic measurability and boundedness requirements, so that $\tilde f_t=\bar\alpha^{t-1}\cdot \tilde f$ for $ t=1,2,....$. Let us use $\bar f$ for the bound $\bar f_1$ that appeared in~(\ref{d-payoff}). In addition, there is a state transition kernel $\tilde g\in {\cal G}(S,X)$, so that $\tilde g_t=\tilde g$ for $t=1,2,...$. For $\chi\in {\cal K}(S,X)$, denote by $T(\chi)$ the operator on ${\cal P}(S)$, so that for any $\sigma\in {\cal P}(S)$,
\begin{equation}\label{T-def-s}
T(\chi)\circ \sigma 
=\sigma\odot\chi\odot\tilde g(\cdot,\cdot,\sigma\otimes\chi).
\end{equation}
Thus, state transition has been made stationary by the stationarity of $\tilde g$.

Denote the stationary nonatomic game formed from the above $S$, $X$, $\bar\alpha$, $\tilde f$, and $\tilde g$ by $\Gamma^\infty$. It helps to first study the corresponding games $\Gamma^t$ that terminate in periods $t+1$, for $t=0,1,...$.
Now let $v^t(s,\xi_{[1t]},\sigma,\chi_{[1t]})$ be the total expected payoff a player can receive in game $\Gamma^t$, when he starts at state $s\in S$ in period 1 and adopts action plan $\xi_{[1t]}\in ({\cal K}(S,X))^t$ from period 1 to $t$, while all other players form state distribution $\sigma\in {\cal P}(S)$ in the beginning and act according to $\chi_{[1t]}\in ({\cal K}(S,X))^t$ from period 1 to $t$. As a terminal condition, we have $v^0(s,\sigma)=0$. Also, for $t=1,2,...$,
\begin{equation}\label{recursive-s}\begin{array}{l}
v^t(s,\xi_{[1t]},\sigma,\chi_{[1t]})=\int_X \xi_1(s|dx)\cdot[\tilde f(s,x,\sigma\otimes \chi_1)\\
\;\;\;\;\;\;\;\;\;\;\;\;+\bar\alpha\cdot\int_S \tilde g(s,x,\sigma\otimes \chi_1|ds')\cdot v^{t-1}(s',\xi_{[2t]},T(\chi_1)\circ\sigma,\chi_{[2t]})].
\end{array}\end{equation}

Using the terminal condition and~(\ref{recursive-s}), we can inductively show that
\begin{equation}\label{bounded-s}
\mid v^{t+1}(s,\xi_{[1,t+1]},\sigma,\chi_{[1,t+1]})-v^t(s,\xi_{[1t]},\sigma,\chi_{[1t]})\mid\leq \bar\alpha^t\cdot\bar f.
\end{equation}
Given $s\in S$, $\xi_{[1\infty]}=(\xi_1,\xi_2,...)\in ({\cal K}(S,X))^\infty$, $\sigma\in {\cal P}(S)$, and $\chi_{[1\infty]}=(\chi_1,\chi_2,...)\in ({\cal K}(S,X))^\infty$, the sequence $\{v^t(s,\xi_{[1t]},\sigma,\chi_{[1t]})\mid t=0,1,...\}$ is thus Cauchy and has a limit point $v^\infty(s,\xi_{[1\infty]},\sigma,\chi_{[1\infty]})$. The latter is the total discounted expected payoff a player can obtain in the game $\Gamma^\infty$, when he starts at state $s$ and adopts action plan $\xi_{[1\infty]}$, while all other players form initial pre-action environment $\sigma$ and act according to $\chi_{[1\infty]}$.

A pre-action environment $\sigma\in {\cal P}(S)$ is said to be associated with $\chi\in {\cal K}(S,X)$ when
\begin{equation}\label{compatible-s}
\sigma=T(\chi)\circ \sigma.
\end{equation}
That is, we let environment $\sigma$ be associated with action plan $\chi$ when the former is invariant under the one-period transition when all players adhere to the latter. For $\chi\in {\cal K}(S,X)$, we use $\chi^\infty$ to represent the stationary policy profile $(\chi,\chi,...)\in ({\cal K}(S,X))^\infty$ that players are to adopt in all periods $t=1,2,...$.

We deem one-time action plan $\chi\in {\cal K}(S,X)$ a stationary Markov equilibrium for the nonatomic game $\Gamma^\infty$, when there exists a $\sigma\in {\cal P}(S)$ that is associated with the given $\chi$, so that for every one-time unilateral deviation $\xi\in {\cal K}(S,X)$,
\begin{equation}\label{first-s}
v^\infty(s,\chi^\infty,\sigma,\chi^\infty) \geq v^\infty(s,(\xi,\chi^\infty),\sigma,\chi^\infty),\hspace*{.5in}\forall s\in S.
\end{equation}
Therefore, a policy will be considered an equilibrium when it induces an invariant environment under whose sway the policy turns out to be a best response in the long run.

Now we move on to the $n$-player game $\Gamma^\infty_n$ made out of the same $S$, $X$, $\bar\alpha$, $\tilde f$, and $\tilde g$. Similarly to the above, we let $\Gamma^t_n$ be its $n$-player counterpart that terminates in period $t+1$. Now let $v^t_n(s_1,\xi_{[1t]},\varepsilon_{s_{-1}},\chi_{[1t]})$ be the total expected payoff player 1 can receive in game $\Gamma^n_t$, when he starts with state $s_1\in S$ and adopts action plan $\xi_{[1t]}\in ({\cal K}(S,X))^t$ from period 1 to $t$, while other players form initial empirical distribution $\varepsilon_{s_{-1}}=\varepsilon_{(s_2,...,s_n)}\in {\cal P}_{n-1}(S)$ and adopt policy $\chi_{[1t]}\in ({\cal K}(S,X))^t$ from 1 to $t$. As a terminal condition, we have $v^0_n(s_1,\varepsilon_{s_{-1}})=0$. For $t=1,2,...$, it follows that
\begin{equation}\label{recursive-n-s}\begin{array}{l}
v^t_n(s_1,\xi_{[1t]},\varepsilon_{s_{-1}},\chi_{[1t]})=\int_X\xi_1(s_1|dx_1)\cdot \int_{X^{n-1}}\chi_1^{\; n-1}(s_{-1}|dx_{-1})\cdot[\tilde f(s_1,x_1,\varepsilon_{s_{-1}x_{-1}})\\
\;\;\;\;\;\;\;\;\;\;\;\;\;\;\;\;\;\;\;\;\;\;\;\;\;\;\;\;\;\;+\bar\alpha\cdot\int_{S^n}\tilde g^n(s,x|ds')\cdot v^{t-1}_n(s'_1,\xi_{[2t]},\varepsilon_{s'_{-1}},\chi_{[2t]})].
\end{array}\end{equation}

Using the terminal condition and~(\ref{recursive-n-s}), we can inductively show that
\begin{equation}\label{bounded-ss}
\mid v^{t+1}_n(s_1,\xi_{[1,t+1]},\varepsilon_{s_{-1}},\chi_{[1,t+1]})-v^t_n(s_1,\xi_{[1t]},\varepsilon_{s_{-1}},\chi_{[1t]})\mid\leq \bar\alpha^t\cdot\bar f.
\end{equation}
Given $s_1\in S$, $\xi_{[1\infty]}\in ({\cal K}(S,X))^\infty$, $\varepsilon_{s_{-1}}\in {\cal P}_{n-1}(S)$, and $\chi_{[1\infty]}\in ({\cal K}(S,X))^\infty$, the sequence $\{v^n_t(s_1,\xi_{[1t]},\varepsilon_{s_{-1}},\chi_{[1t]})\mid t=0,1,...\}$ is Cauchy and has a limit point $v^\infty_n(s_1,\xi_{[1\infty]},\varepsilon_{s_{-1}},\chi_{[1\infty]})$. The latter is the total discounted expected payoff a player can obtain in $\Gamma^\infty_n$, when he starts at state $s$ and adopts action plan $\xi_{[1\infty]}$, while all other players form the initial pre-action environment $\varepsilon_{s_{-1}}$ and act according to $\chi_{[1\infty]}$.

For the current setting, it should be noted that Assumptions~\ref{g-c} and~\ref{f-c} translate into the continuity in $\tau$ at an $(s,x)$-independent rate of, respectively, the transition kernel $\tilde g(s,x,\tau)$ and payoff function $\tilde f(s,x,\tau)$. We now present the main result for the stationary case.

\begin{theorem}\label{main-s}
Suppose $\chi\in {\cal K}(S,X)$ is a stationary Markov equilibrium for the stationary nonatomic game $\Gamma^\infty$. Let $\hat\pi_{n-1}\in {\cal P}(S^{n-1})$ for each $n\in\mathbb{N}\setminus\{1\}$. Also suppose the sequence $\hat\pi_{n-1}$ asymptotically resembles the sequence $\sigma^{n-1}$, where $\sigma$ is associated with $\chi$ in the equilibrium definitions~(\ref{compatible-s}) and~(\ref{first-s}). Then, $\chi^\infty$ would be asymptotically equilibrium for games $\Gamma^\infty_n$ in an average sense. More specifically, for any $\epsilon>0$ and large enough $n\in \mathbb{N}$,
\[\int_{S^{n-1}}\hat\pi_{n-1}(ds_{-1})\cdot v^\infty_n(s_1,\chi^\infty,\varepsilon_{s_{-1}},\chi^\infty)\geq \int_{S^{n-1}}\hat\pi_{n-1}(ds_{-1})\cdot  v^\infty_n(s_1,\xi_{[1\infty]},\varepsilon_{s_{-1}},\chi^\infty)-\epsilon,\]
for any $s_1\in S$ and $\xi_{[1\infty]}\in ({\cal K}(S,X))^\infty$.
\end{theorem}


Theorem~\ref{main-s} says that, players in a large finite stationary game will not regret much by adopting a stationary equilibrium for a correspondent stationary nonatomic game. The regret can be measured in an average sense, so long as the underlying other-player multi-state distribution $\hat\pi_{n-1}$ is close to an invariant $\sigma$ associated with the NG equilibrium. 
Just as in Section~\ref{interpretation}, we can let $\hat\pi_{n-1}=\sigma^{n-1}$, indicating that players take a ``lazy'' approach in assessing other players' states. We leave discussion of other possibilities to Appendx~\ref{app-e}.

\section{Implications of Main Results}\label{existence}


\subsection{Observation, Remembrance, and Coordination}

Regarding Theorems~\ref{main} and~\ref{main-f}, we note the following for $\bar t$-period games. A prominent feature of an NG equilibrium $\chi_{[1\bar t]}\in ({\cal K}(S,X))^{\bar t}$ is its insensitivity, at any period $t$, to a player's personal history $(s_{t'},x_{t'}|t'=1,2,...,t-1)$, historical data regarding other players, and the present information about other players' states. Independence of the first two factors has much to do with the Markovian setup of the game---neither $\tilde f_t$ nor $\tilde g_t$ depends on past history. But the more interesting independence of the latter two factors stems from players' common knowledge about the evolution of their environments. 
The $(\sigma_{t'}\otimes \chi_{t'}|t'=1,2,...,t-1)$ portion of the history 
and the present information $\sigma_t$, both about other players, are determinable by~(\ref{sequence}) before the game is even played out.

For finite semi-anonymous games, however, information is gradually revealed and its perfection is not guaranteed. We can define space $O_S$ and map $\tilde o_S:{\cal P}(S)\rightarrow O_S$ to represent a player's observatory power over his present pre-action environment immediately before actual play. Similarly, we can define space $O_{SX}$ and map $\tilde o_{SX}:{\cal P}(S\times X)\rightarrow O_{SX}$ to represent his observatory power over the in-action environment just experienced. So that new information does not contradict old information and no information gets lost, we suppose function $\tilde o_S^{\;SX}:O_{SX}\rightarrow O_S$ exists, with $\tilde o_S^{\;SX}(\tilde o_{SX}(\tau))=\tilde o_S(\tau|_S)$ for any $\tau\in {\cal P}(S\times X)$.

With these definitions, a player's decision in period $t$ can be denoted by a map $\hat\chi_t: (S\times X\times O_{SX})^{t-1}\times O_S\times S\rightarrow {\cal P}(X)$. In the period, player 1's random decision rule can be written as $\hat\chi_t(\tilde h_t,\tilde o_S(\varepsilon_{s_{t,-1}}),s_{t1}|\cdot)$, where the history $\tilde h_t$ is expressible as
\begin{equation}
\tilde h_t=(s_{t'1},x_{t'1},\tilde o_{SX}(\varepsilon_{s_{t',-1}x_{t',-1}})|t'=1,2,...,t-1),
\end{equation}
$\tilde o_S(\varepsilon_{s_{t,-1}})$ is his observation of other players' status, and $s_{t1}$ represents the player's own state. There is a whole spectrum in which $O_S$ and $\tilde o_S$ can reside. When $O_S=\{0\}$ and $\tilde o_S(\cdot)=0$, players are ignorant of others' states; when $O_S={\cal P}(S)$ and $\tilde o_S$ is the identity map, every player is fully aware of his surrounding. Similarly, there are varieties of $O_{SX}$, $\tilde o_{SX}$, and $\tilde o_S^{\;SX}$.

Theorems~\ref{main} and~\ref{main-f}, however, nullify the need to delve into the $(O_S,\tilde o_S,O_{SX},\tilde o_{SX},\tilde o_S^{\;SX})$-related details about finite games. They state that an equilibrium of the NG counterpart, which is necessarily both oblivious of the past history $\tilde h_t$ and blind to the present observation $\tilde o_S(\varepsilon_{s_{t,-1}})$, serves as a good approximate equilibrium for games with enough players.
The absence of $\tilde h_t$ again has a Markovian explanation.  On the other hand, the ability to shake off $\tilde o_S(\varepsilon_{s_{t,-1}})$'s influence is very important, since this saves players the efforts to gather information about their surroundings. 

Regarding Theorem~\ref{main-s}, we note the following. Each of our finite stationary games is a discounted stochastic game.
For an $n$-player version of the latter game in which players have full knowledge of others' states, equilibria are hard to compute and for their implementation, require high degrees of coordination among players; see Solan \cite{S98}. These equilibria come from the space $(2^{\mathbb{R}^n})^{S^n}\times ((\mathbb{R}^n)^{X^n\times S^n})^{S^n\times \mathbb{R}^n}$; whereas, our NG equilibria come from $\mathbb{R}^{S\times X}$. Meanwhile, the discounted stochastic game one faces in real life is often semi-anonymous; see, e.g., examples listed in Jovanoic and Rosenthal \cite{JR88}. For such a game, Theorem~\ref{main-s} has shown that a much easier path can be taken in order to coordinate player behavior under an $\epsilon$-sized compromise. If players all agree to exercise a corresponding NG equilibrium, the typical player 1 has only to respond to his own state $s_{t1}$ without giving up too much.

\subsection{Sources of NG Equilibria}

To further buttress the claim that studying the idealistic NGs can help with the understanding and execution of messier finite games faced in real life, we demonstrate that NG equilibria, meeting criteria~(\ref{first}) and~(\ref{environ}) for the transient case and~(\ref{compatible-s}) and~(\ref{first-s}) for the stationary case, can be obtained relatively easily. 

First, we concentrate on the transient case studied in Sections~\ref{overture} to~\ref{equilibria}. From~(\ref{recursive}),
\begin{equation}
v_t(s_t,(\xi_t,\chi_{[t+1,\bar t]}),\sigma_t,\chi_{[t\bar t]})=\int_X \xi_t(dy)\cdot v_t(s_t,(\delta_y,\chi_{[t+1,\bar t]}),\sigma_t,\chi_{[t\bar t]}).
\end{equation}
Hence,
\begin{equation}
\sup_{\xi_t\in {\cal K}(S,X)}v_t(s_t,(\xi_t,\chi_{[t+1,\bar t]}),\sigma_t,\chi_{[t\bar t]})=\sup_{y\in X}v_t(s_t,(\delta_y,\chi_{[t+1,\bar t]}),\sigma_t,\chi_{[t\bar t]}).
\end{equation}
So the equilibrium criterion~(\ref{first}) conveniently used by us for the $\bar t$-period case is equivalent to, for every $t=1,2,...,\bar t$,
\begin{equation}\label{first-alt}
\chi_t(s_t|\tilde X_t(s_t,\sigma_t,\chi_{[t\bar t]}))=1,\hspace*{.5in}\forall s_t\in S,
\end{equation}
where
\begin{equation}\label{xx-def}
\tilde X_t(s_t,\sigma_t,\chi_{[t\bar t]})=\{x\in X|v_t(s_t,(\delta_x,\chi_{[t+1,\bar t]}),\sigma_t,\chi_{[t\bar t]})=\sup_{y\in X}v_t(s_t,(\delta_y,\chi_{[t+1,\bar t]}),\sigma_t,\chi_{[t\bar t]})\},
\end{equation}
and $\sigma_t$ is defined through~(\ref{environ}). 

The form consisting of~(\ref{first-alt}) and~(\ref{xx-def}) is fairly close to the distributional-equilibrium concept used in NG literature, such as Mas-Colell \cite{M84} and Jovanovic and Rosenthal \cite{JR88}. A distributional equilibrium is an in-action environment sequence $\tau_{[1\bar t]}=(\tau_t|t=1,2,...,\bar t)\in ({\cal P}(S\times X))^{\bar t}$ which satisfies $\tau_t(\tilde U_t(\tau_{[t\bar t]}))=1$ for each $t=1,2,...,\bar t$. Here, $\tilde U_t(\tau_{[t\bar t]})=
\{(s,x)\in S\times X|v'_t(s,x,\tau_{[t\bar t]})=\sup_{y\in X}v'_t(s,y,\tau_{[t\bar t]})\}$, and $v'_t(s,y,\tau_{[t\bar t]})$ is a player's payoff when he starts period $t$ with state $s$ and action $y$, but other players in all periods and he himself in later periods act according to $\tau_{[t\bar t]}$; corresponding to~(\ref{environ}), the distributional equilibrium also satisfies $\tau_1|_S=\sigma_1$ and $\tau_t|_S=\tau_{t-1}\odot \tilde g_{t-1}(\cdot,\cdot,\tau_{t-1})$ for $t=2,3,...,\bar t$. According to Jovanovic and Rosenthal \cite{JR88} (Theorem 1), such an equilibrium $\tau_{[1\bar t]}$ would exist when $S$ and $X$ are compact, each payoff $\tilde f_t$ is bounded and continuous in all arguments, and each transition kernel $\tilde g_t$ is continuous in all arguments.

When an equilibrium $\chi_{[1\bar t]}$ in our conditional sense exists, we can construct a distributional equilibrium $\tau_{[1\bar t]}$ by resorting iteratively to $\tau_t=\sigma_t\otimes\chi_t$ and $\sigma_{t+1}=T_t(\chi_t)\circ\sigma_t$ for $t=1,2,...,\bar t$. 
Conversely, when the latter distributional equilibrium $\tau_{[1\bar t]}$ is available, we can nearly get a conditional equilibrium $\chi_{[1\bar t]}$ back. For each $t=1,2,...,\bar t$, according to Duffie, Geanakoplos, Mas-Colell, and McLennan \cite{DGMM94} (p. 751), we can identify a $\chi_t\in {\cal K}(S,X)$, which also passes as a measurable map from $S$ to ${\cal P}(X)$, that satisfies $\tau_t=\tau_t|_S\otimes\chi_t$. Thus, we will be able to construct $\chi_{[1\bar t]}$ consecutively from $\chi_1$ up to $\chi_{\bar t}$. But even then, $\chi_{[t\bar t]}$ along with $\sigma_t=\tau_t|_S$ would satisfy~(\ref{first-alt}) only for $\tau_t|_S$-almost every $s_t$, but not necessarily every $s_t\in S$. For instance, 
we can suppose $S=\{\bar s_1,\bar s_2,...\}$. At each $t$, the constructed $\chi_{[1\bar t]}$ could guarantee~(\ref{first-alt}) for those $\bar s_i$'s with $(\tau_t|_S)(\bar s_i)>0$ but not those with $(\tau_t|_S)(\bar s_i)=0$. On the other hand, a conditional equilibrium $\chi_{[1\bar t]}$ can be obtained directly; see Appendix~\ref{app-f-t} for details.

When it comes to the stationary case examined in Section~\ref{stationary}, we make parallel developments. Here the property corresponding to~(\ref{first-s}) is
\begin{equation}\label{first-alt-s}
\chi(s|\tilde X_\infty(s,\sigma,\chi))=1,\hspace*{.5in}\forall s\in S,
\end{equation}
where
\begin{equation}\label{xx-def-s}
\tilde X_\infty(s,\sigma,\chi)=\{x\in X|v^\infty(s,(\delta_x,\chi^\infty),\sigma,\chi^\infty)=\sup_{y\in X}v^\infty(s,(\delta_y,\chi^\infty),\sigma,\chi^\infty)\},
\end{equation}
and $\sigma$ satisfies~(\ref{compatible-s}). Again, the existence of a related distributional equilibrium $\tau\in {\cal P}(S\times X)$ is known under quite general conditions; see, e.g., Jovanovic and Rosenthal \cite{JR88} (Theorem 2). However, an equilibrium $\tau$ does not exactly lead to a conditional equilibrium $\chi$. So once more we focus on a direct approach for the stationary case; see Appendix~\ref{app-f-s}.

\section{Concluding Remarks}\label{discussion}

Under a common action plan, we have shown that environments faced by players in multi-period large finite games would stay close to those of their NG counterparts. For transient and stationary settings, our results reveal that an NG equilibrium, necessarily both oblivious of past history and blind to present status of other players, could serve as a good approximate equilibrium in large finite games. 
We reckon that the discreteness requirement on both the state and action spaces can be frustrating in some circumstances. Besides the relaxation of the aforementioned restriction, future research can also look into the issue of converge rate. 
\\ \\

\noindent{\bf\Large Acknowledgments}

This research was supported in part by National Science Foundation Grant CMMI-0854803, as well as National Natural Science Foundation of China Grants 11371273 and 71502015. \\ \\

\noindent{\bf\Large Appendices}\appendix
\numberwithin{equation}{section}

\section{Concepts and Rudimentary Lemmas}\label{app-a}

Recall that $\rho_A$ stands for the Prohorov metric for the space of distributions ${\cal P}(A)$. 

\begin{lemma}\label{newlemma}
	Let $A$ be a separable metric space. Then, for any $n\in \mathbb{N}$ and $a,a'\in A^n$,
	\[ \rho_A(\varepsilon_a,\varepsilon_{a'})\leq\max_{m=1}^n d_A(a_m,a'_m). \]
\end{lemma}
\noindent{\bf Proof: }Let $\epsilon=\max_{m=1}^n d_A(a_m,a'_m)$. For any $A'\in {\cal B}(A)$, the key observation is that
\begin{equation}
\delta_{a'_m}((A')^\epsilon)\geq \delta_{a_m}(A').
\end{equation}
Then,
\begin{equation}
\varepsilon_{a'}((A')^\epsilon)=\frac{\sum_{m=1}^n\delta_{a'_m}((A')^\epsilon)}{n}\geq \frac{\sum_{m=1}^n\delta_{a_m}(A')}{n}=\varepsilon_a(A').
\end{equation}
Thus, $\rho_A(\varepsilon_a,\varepsilon_{a'})\leq \epsilon$. \qed

According to Parthasarathy \cite{P05} (Theorem II.7.1), the strong law of large numbers applies to the empirical distribution under the weak topology, and hence under the Prohorov metric. In the following, we state its weak version.

\begin{lemma}\label{p-Prohorov}
Let separable metric space $A$ and distribution $p\in {\cal P}(A)$ be given. Then, for any $\epsilon>0$, as long as $n$ is large enough,
\[ p^n(\{a\in A^n\mid \rho_A(\varepsilon_a ,p )<\epsilon\})>1-\epsilon. \]
\end{lemma}

Due to the inequality of Dvoretzky, Kiefer, and Wolfolwitz \cite{DKW56}, the above convergence is uniform for certain $A$'s. The inequality infers that, when $A$ is $\mathbb{R}$ or countable,
\[
p^n(\{a\in A^n\mid \rho_A(\varepsilon_a,p)\leq \epsilon\})>1-2e^{-2n\epsilon^2},\hspace*{.5in}\forall \epsilon>0.
\]
When $n$ is greater than $\ln(3/\epsilon)/(2\epsilon^2)$, a number independent of $p\in {\cal P}(A)$, the above would entail the inequality in Lemma~\ref{p-Prohorov}. Thus, we have the following.

\begin{lemma}\label{p-uniform}
When $A$ is the real line $\mathbb{R}$ or countable, the convergence expressed in Lemma~\ref{p-Prohorov} is uniform. Namely, a lower bound could be identified so that every $n$ above it would realize the inequality in the lemma for every $p\in {\cal P}(A)$.
\end{lemma}

For separable metric space $A$, point $a\in A$, and the $(n-1)$-point empirical distribution $p\in {\cal P}_{n-1}(A)$, we use $(a,p)_n$ to represent the member of ${\cal P}_n(A)$ that has an additional $1/n$ weight on the point $a$, but with probability masses in $p$ being reduced to $(n-1)/n$ times of their original values. For $a\in A^n$ and $m=1,...,n$, we have $(a_m,\varepsilon_{a_{-m}})_n=\varepsilon_a$. Concerning the Prohorov metric, we have also a simple but useful observation.

\begin{lemma}\label{p-mc8}
Let $A$ be a separable metric space. Then, for any $n\in \mathbb{N}\setminus\{1\}$, $a\in A$, and $p\in {\cal P}_{n-1}(A)$,
\[\rho_A((a,p)_n,p)\leq \frac{1}{n}. \]
\end{lemma}
\noindent{\bf Proof: }
Let $A'\in {\cal B}(A)$ be chosen. Then $p(A')=(m-1)/(n-1)$ for some $m=1,2,...,n$. If $a\notin A'$, then $(a,p)_n(A')=(m-1)/n$ and hence
\begin{equation}
(a,p)_n(A')\leq p(A')\leq (a,p)_n(A')+\frac{1}{n}.
\end{equation}
If $a\in A'$, then $(a,p)_n(A')=m/n$ and hence
\begin{equation}
(a,p)_n(A')-\frac{1}{n}\leq p(A')\leq (a,p)_n(A').
\end{equation}
Therefore, it is always true that
\begin{equation}
\mid (a,p)_n(A')-p(A')\mid\leq \frac{1}{n}.
\end{equation}
Due to the nature of the Prohorov metric, we have
\begin{equation}
\rho_A((a,p)_n,p)\leq \frac{1}{n}.
\end{equation}
We have thus completed the proof.\qed

For the notion of asymptotic resemblance introduced in Definition~\ref{conv-p}, we have that it is preserved under certain projections and expansions.

\begin{lemma}\label{d-prob}
Let $A$ be a separable metric space. Also, $q_n\in {\cal P}(A^n)$ for every $n\in \mathbb{N}$ and $p\in {\cal P}(A)$. Suppose the sequence $q_n$ asymptotically resembles the sequence $p^n$. Then, the sequence $q_n|_{A^{n-1}}$ will asymptotically resemble the sequence $p^{n-1}$.
\end{lemma}
\noindent{\bf Proof: }For any $\epsilon>0$, due to the asymptotic resemblance of the sequence $q_n$ to the sequence $p^n$, we have, for $n$ large enough,
\begin{equation}\label{a78a}
q_n(A'_n)>1-\epsilon,
\end{equation}
where
\begin{equation}\label{b78a}
A'_n=\{a\in A^n|\rho_A(\varepsilon_a,p)<\epsilon\}.
\end{equation}
By Lemma~\ref{p-mc8}, we have
\begin{equation}
\rho_A(\varepsilon_a,\varepsilon_{a_{-1}})\leq\frac{1}{n},\hspace*{.5in}\forall a\in A^n.
\end{equation}
Hence, for large enough $n$,
\begin{equation}\label{c78a}
A'_n\subseteq A\times A''_{n-1},
\end{equation}
where
\begin{equation}\label{d78a}
A''_{n-1}=\{a_{-1}\in A^{n-1}|\rho_A(\varepsilon_{a_{-1}},p)<2\epsilon\}.
\end{equation}
But by~(\ref{a78a}), this means that
\begin{equation}
(q_n|_{A^{n-1}})(A''_{n-1})=q_n(A\times A''_{n-1})\geq q_n(A'_n)>1-\epsilon.
\end{equation}
That is, $q_n|_{A^{n-1}}$ asymptotically resembles $p^{n-1}$. \qed

\begin{lemma}\label{dc-prob}
Let $A$ be a separable metric space. Also, $q_n\in {\cal P}(A^n)$ for every $n\in \mathbb{N}$ and $p,p'\in {\cal P}(A)$. Suppose the sequence $q_n$ asymptotically resembles the sequence $p^n$. Then, the sequence $p'\times q_{n-1}$ will asymptotically resemble the sequence $p^n$ as well.
\end{lemma}
\noindent{\bf Proof: }For any $\epsilon>0$, due to the asymptotic resemblance of the sequence $q_n$ to the sequence $p^n$, we have, for $n$ large enough,
\begin{equation}\label{a78a-n-1}
q_{n-1}(A'_{n-1})>1-\epsilon,
\end{equation}
where
\begin{equation}\label{b78a-n-1}
A'_{n-1}=\{a\in A^{n-1}|\rho_A(\varepsilon_a,p)<\epsilon\}.
\end{equation}
By Lemma~\ref{p-mc8}, we have
\begin{equation}
\rho_A(\varepsilon_{(a_1,a)},\varepsilon_a)\leq\frac{1}{n},\hspace*{.5in}\forall a_1\in A,\;a\in A^{n-1}.
\end{equation}
Hence, for large enough $n$,
\begin{equation}\label{c78a-n-1}
A\times A'_{n-1}\subseteq A''_n,
\end{equation}
where
\begin{equation}\label{d78a-n-1}
A''_n=\{a\in A^n|\rho_A(\varepsilon_a,p)<2\epsilon\}.
\end{equation}
But by~(\ref{a78a-n-1}), this means that
\begin{equation}
(p'\times q_{n-1})(A''_n)\geq (p'\times q_{n-1})(A\times A'_{n-1})=p'(A)\times q_{n-1}(A'_{n-1})>1-\epsilon.
\end{equation}
That is, $p'\times q_{n-1}$ asymptotically resembles $p^n$. \qed

\begin{lemma}\label{c-prob}
Let $A$ and $B$ be separable metric spaces. Also, $q_n\in {\cal P}(A^n\times B^n)$ for every $n\in \mathbb{N}$ and $p\in {\cal P}(A\times B)$. Suppose the sequence $q_n$ asymptotically resembles the sequence $p^n$. Then, the sequence $q_n|_{A^n}$ will asymptotically resemble the sequence $(p|_A)^n$.
\end{lemma}
\noindent{\bf Proof: }For any $\epsilon>0$, due to the asymptotic resemblance of the sequence $q_n$ to the sequence $p^n$, we have, for $n$ large enough,
\begin{equation}\label{a78}
q_n(C'_n)>1-\epsilon,
\end{equation}
where 
\begin{equation}\label{b78}
C'_n=\{c=(a,b)\in A^n\times B^n|\rho_{A\times B}(\varepsilon_c,p)<\epsilon\}.
\end{equation}
But by (87) of Yang \cite{Y11},
\begin{equation}
\rho_A(\varepsilon_a,p|_A)=\rho_A(\varepsilon_c|_A,p|_A)\leq \rho_{A\times B}(\varepsilon_c,p),\hspace*{.5in}\forall c=(a,b)\in C'_n.
\end{equation}
Hence,
\begin{equation}\label{c78}
C'_n\subseteq A'_n\times B^n,
\end{equation}
where
\begin{equation}\label{d78}
A'_n=\{a\in A^n|\rho_A(\varepsilon_a,p|_A)<\epsilon\}.
\end{equation}
Combining~(\ref{a78}) and~(\ref{c78}), we can obtain
\begin{equation}
(q_n|_{A^n})(A'_n)=q_n(A'_n\times B^n)\geq q_n(C'_n)>1-\epsilon.
\end{equation}
This indicates that $q_n|_{A^n}$ asymptotically resembles $(p|_A)^n$.\qed

\section{Proofs of Section~\ref{convergence}}\label{app-b}

\noindent{\bf Proof of Proposition~\ref{T-onestep}: }We first prove (i). Fix some $\epsilon\in (0,1)$. Due to the countability of $S$
, we can identify some $I$ of its points $\bar s_1,\bar s_2,...,\bar s_I$, so that each $\sigma(\{\bar s_i\})>0$ and
\begin{equation}\label{usedlater}
\sum_{i=1}^I\sigma(\{\bar s_i\})>1-\epsilon.
\end{equation}
For convenience, let $\bar S'=\{\bar s_1,\bar s_2,...,\bar s_I\}$ and $\bar S''=S\setminus\bar S'$.

Since $S$ is discrete, the distance $d_S(\bar S',\bar S'')=\inf_{s'\in \bar S',s''\in \bar S''}d_S(s',s'')>0$. For $i,j=1,2,...,I$, let us use $d_{ij}$ for $d_S(\bar s_i,\bar s_j)$ and $\sigma_i$ for $\sigma(\{\bar s_i\})$. Now define
\begin{equation}\label{delta-def}
\delta=\frac{\epsilon}{I}\wedge d_S(\bar S',\bar S'')\wedge(\min_{i\neq j}d_{ij})\wedge (\min_i\frac{\sigma_i}{2}),
\end{equation}
which is still strictly positive. In this paper, we use $a\wedge b$ to stand for $\min\{a,b\}$ and $a\vee b$ to stand for $\max\{a,b\}$.

For any $n\in \mathbb{N}$, define $S'_n\in {\cal B}(S^n)$ so that
\begin{equation}\label{new-d}
S'_n=\{s\in S^n|\rho_S(\varepsilon_s,\sigma)<\delta\}.
\end{equation}
By the hypothesis that $\pi_n$ asymptotically resembles $\sigma^n$, we can ensure
\begin{equation}\label{big1}
\pi_n(S'_n)>1-\frac{\epsilon}{2},
\end{equation}
by making $n$ large enough.

Consider any such $n$, as well as any $s=(s_1,s_2,...,s_n)\in S'_n$ and $i=1,2,...,I$. It follows from $\delta\leq d_S(\bar S',\bar S'')\wedge(\min_{i\neq j}d_{ij})$ that $(\{\bar s_i\})^\delta$, whose meaning comes from~(\ref{qm0}), is still $\{\bar s_i\}$ itself. Now by~(\ref{new-d}),
\begin{equation}\label{later-d}
\varepsilon_s(\{\bar s_i\})<\sigma((\{\bar s_i\})^\delta)+\delta=\sigma_i+\delta,
\end{equation}
and
\begin{equation}\label{evenlater}\begin{array}{l}
\varepsilon_s(\{\bar s_i\})=1-\varepsilon_s(\{\bar s_j|j\neq i\}\cup \bar S'')>1-\sigma((\{\bar s_j|j\neq i\}\cup \bar S'')^\delta)-\delta\\
\;\;\;\;\;\;\;\;\;\;\;\;=1-\sigma(\{\bar s_j|j\neq i\}\cup \bar S'')-\delta=\sigma_i-\delta,
\end{array}\end{equation}
which is still above $\delta>0$ by the fact that $\delta\leq \min_i\sigma_i/2$. For convenience, let $n_i(s)=n\cdot \varepsilon_s(\{\bar s_i\})$, the number of components $s_m$ of $s$ that happen to be $\bar s_i$. Now we know that $n_i(s)$ is above $n\delta$ for every $s\in S'_n$ and $i=1,2,...,I$. 

On the other hand, by Lemma~\ref{p-Prohorov}, there exists some $\underline n_i$ for each $i=1,2,...,I$, so that when $n_i>\underline n_i$,
\begin{equation}\label{big2}
(\chi(\bar s_i))^{n_i}(X'_{in_i})>1-\frac{\epsilon}{2I},
\end{equation}
where
\begin{equation}\label{big3}
X'_{in_i}=\{x\in X^{n_i}|\rho_X(\varepsilon_x,\chi(\bar s_i))<\delta\}.
\end{equation}
Since $\delta>0$, we can ensure that $n\delta$ and hence $n_i(s)$ is above $\underline n_i$ for every $i=1,2,...,I$ by letting $n$ be large enough.

Fix a big $n$ that facilitates both~(\ref{big1}) and~(\ref{big2}). For any $(s,x)\in S^n\times X^n$, let $\tilde x_i(s,x)$ be the $n_i(s)$-long vector of $x_m$'s whose corresponding $s_m$'s happen to be $\bar s_i$:
\begin{equation}
\tilde x_i(s,x)=(x_m|m=1,2,...,n\mbox{ but with }s_m=\bar s_i)\in X^{n_i(s)}.
\end{equation}
Define $U'_n\in {\cal B}(S^n\times X^n)$, so that
\begin{equation}
U'_n=\{(s,x)\in S^n\times X^n|s\in S'_n\mbox{ and }\tilde x_i(s,x)\in X'_{in_i(s)}\mbox{ for each }i=1,2,...,I\}.
\end{equation}
By~(\ref{consult3}),~(\ref{big1}), and~(\ref{big2}), we have
\begin{equation}\label{afact}
(\pi_n\otimes\chi^n)(U'_n)=\int_{S'_n}\pi_n(ds)\cdot\prod_{i=1}^I(\chi(\bar s_i))^{n_i(s)}(X'_{in_i(s)})>(1-\frac{\epsilon}{2})\cdot (1-\frac{\epsilon}{2I})^I>1-\epsilon.
\end{equation}

For any $(s,x)$ in $U'_n$, let us examine how close $\varepsilon_{sx}=\varepsilon_{((s_1,x_1),...,(s_n,x_n))}$ is to $\sigma\otimes \chi$. Recall that $S=\{\bar s_1,\bar s_2,...,\bar s_I\}\cup\bar S''$. So for any $U'\in {\cal B}(S\times X)$,
\begin{equation}\label{aaa}
U'=(\bigcup_{i=1}^I\{\bar s_i\}\times X'_i)\bigcup U'',
\end{equation}
where $X'_i\in {\cal B}(X)$ for $i=1,2,...,I$, while $U''$ is such that $s''\in \bar S''$ for any $(s'',x'')\in U''$. Note again that $\delta\leq d_S(\bar S',\bar S'')\wedge \min_{i\neq j}d_{ij}$. When we take $d_{S\times X}$ to mean $d_{S\times X}((s',x'),(s'',x''))=d_S(s',s'')\vee d_X(x',x'')$,~(\ref{aaa}) would lead to
\begin{equation}\label{a2a}
\bigcup_{i=1}^I\{\bar s_i\}\times (X'_i)^\delta \subseteq (U')^\delta.
\end{equation}

Now from~(\ref{later-d}) and~(\ref{big3}),
\begin{equation}\begin{array}{l}
\varepsilon_{sx}(\{\bar s_i\}\times X'_i)=\varepsilon_s(\{\bar s_i\})\cdot \varepsilon_{\tilde x_i(s,x)}(X'_i)<(\sigma_i+\delta)\cdot [\chi(\bar s_i|(X'_i)^\delta)+\delta]\\
\;\;\;\;\;\;\leq(\sigma\otimes\chi)(\{\bar s_i\}\times (X'_i)^\delta)+2\delta+\delta^2<(\sigma\otimes\chi)(\{\bar s_i\}\times (X'_i)^\delta)+3\delta,
\end{array}\end{equation}
where the last inequality is due to our choice that $\delta\leq\epsilon/I<1$. Meanwhile,
\begin{equation}\label{bbb}
\varepsilon_{sx}(U'')\leq \varepsilon_{sx}(\bar S''\times X)=\varepsilon_s(\bar S'')=1-\sum_{i=1}^I\varepsilon_s(\{\bar s_i\})<1-\sum_{i=1}^I\sigma_i+I\delta<\epsilon+I\delta,
\end{equation}
where the second-to-last inequality is due to~(\ref{evenlater}) and the last one is due to~(\ref{usedlater}). Combine~(\ref{aaa}) to~(\ref{bbb}), and we can obtain
\begin{equation}
\varepsilon_{sx}(U')<(\sigma\otimes\chi)((U')^\delta)+\epsilon+4I\delta.
\end{equation}
Thus,
\begin{equation}\label{bfact}
\rho_{S\times X}(\varepsilon_{sx},\sigma\otimes\chi)<\epsilon+4I\delta\leq 5\epsilon,
\end{equation}
where the last inequality comes from our choice that $\delta\leq \epsilon/I$. Since~(\ref{afact}) and~(\ref{bfact}) are to occur at any $n$ that is large enough, we see that (i) is true.

We then prove (ii). For convenience, we denote $S\times X$ by $U$, $\sigma\otimes\chi$ by $\tau$, and for each $n\in \mathbb{N}$, $\pi_n\otimes \chi^n$ by $\nu_n$. From (i), we have the sequence $\nu_n$ asymptotically resembling the sequence $\tau^n$.


Fix some $\epsilon\in (0,1)$. Due to the countability of $S$ and $X$ 
, and hence that of $U$, we can identify some $J$ points $\bar u_1,\bar u_2,...,\bar u_J$, so that each $\tau(\{\bar u_j\})>0$ and
\begin{equation}\label{usedlater2}
\sum_{j=1}^J\tau(\{\bar u_j\})>1-\epsilon.
\end{equation}
For convenience, let $\bar U'=\{\bar u_1,\bar u_2,...,\bar u_J\}$ and $\bar U''=U\setminus\bar U'$.

As $S$ and $X$ are both discrete,
 so $U$ is discrete as well. Thence, the distance $d_U(\bar U',\bar U'')=\inf_{u'\in \bar U',u''\in \bar U''}d_U(u',u'')>0$. For $j,k=1,2,...,J$, let us use $d'_{jk}$ for $d_U(\bar u_j,\bar u_k)$ and $\tau_j$ for $\tau(\{\bar u_j\})$. Now define
\begin{equation}\label{delta-def2}
\delta=\frac{\epsilon}{J}\wedge d_U(\bar U',\bar U'')\wedge(\min_{j\neq k}d'_{jk})\wedge (\min_j\frac{\tau_j}{2}),
\end{equation}
which is still strictly positive.

For any $n\in \mathbb{N}$, define $U'_n\in {\cal B}(U^n)$ so that
\begin{equation}\label{new-d2}
U'_n=\{u\in U^n|\rho_U(\varepsilon_u,\tau)\bigvee [2\cdot \sup_{u'\in U}\max_{m=1}^n \rho_S(g(u',\varepsilon_{u_{-m}}),g(u',\tau))]<\delta\}.
\end{equation}
By (i) that $\nu_n$ asymptotically resembles $\tau^n$, the hypothesis that $g(u,\cdot)$ is continuous at a $u$-independent rate, and Lemma~\ref{p-mc8}, we can ensure
\begin{equation}\label{big12}
\nu_n(U'_n)>1-\frac{\epsilon}{2},
\end{equation}
by making $n$ large enough,

Consider any such $n$, as well as any $u=(u_1,u_2,...,u_n)\in U'_n$ and $j=1,2,...,J$. It follows from $\delta\leq d_U(\bar U',\bar U'')\wedge(\min_{j\neq k}d'_{jk})$ that $(\{\bar u_j\})^\delta$ is still $\{\bar u_j\}$ itself. Now by~(\ref{new-d2}),
\begin{equation}\label{later-d2}
\varepsilon_u(\{\bar u_j\})<\tau((\{\bar u_j\})^\delta)+\delta=\tau_j+\delta,
\end{equation}
and
\begin{equation}\label{evenlater2}
\varepsilon_u(\{\bar u_j\})>1-\tau((\{\bar u_k|k\neq j\}\cup \bar U'')^\delta)-\delta=\tau_j-\delta,
\end{equation}
which is still above $\delta>0$ by the fact that $\delta\leq \min_j\tau_j/2$. For convenience, let $n'_j(u)=n\cdot \varepsilon_u(\{\bar u_j\})$. Now we know that $n'_j(u)$ is above $\lfloor n\delta\rfloor$ for every $j=1,2,...,J$.

Due to the countability of $U$ and Lemma~\ref{p-uniform} on the uniform Glivenko-Cantelli property, there exists some $\underline n'$, independent of both $j$ and $u$, such that when every $n'_j(u)>\underline n'$,
\begin{equation}\label{big220}
(g(\bar u_j,\varepsilon_{u\setminus\bar u_j}))^{n'_j(u)}(S''_{jn'_j(u)}(u))>1-\frac{\epsilon}{2J},\hspace*{.5in}\forall j=1,2,...,J,
\end{equation}
where every $u\setminus\bar u_j$ is the $(n-1)$-long vector that is almost identical to $u$ but with only $n'_j(u)-1$ components equal to $\bar u_j$, and
\begin{equation}\label{big320}
S''_{jn'}(u')=\{s\in S^{n'}|\rho_S(\varepsilon_s,g(\bar u_j,\varepsilon_{u'\setminus\bar u_j}))<\frac{\delta}{2}\}.
\end{equation}
But in light of~(\ref{new-d2}), we can really guarantee that
\begin{equation}\label{big22}
(g(\bar u_j,\varepsilon_{u\setminus\bar u_j}))^{n'_j(u)}(S'_{jn'_j(u)})>1-\frac{\epsilon}{2J},\hspace*{.5in}\forall j=1,2,...,J,
\end{equation}
where
\begin{equation}\label{big32}
S'_{jn'}=\{s\in S^{n'}|\rho_S(\varepsilon_s,g(\bar u_j,\tau))<\delta\}.
\end{equation}
Since $\delta>0$, we can ensure that $\lfloor n\delta\rfloor$ and hence $n'_j(u)$ is above $\underline n'$ for every $j=1,2,...,J$ by letting $n$ be large enough.

Fix a big $n$ that facilitates both~(\ref{big12}) and~(\ref{big22}). For any $(u,s)\in U^n\times S^n$, let $\tilde s_j(u,s)$ be the $n'_j(u)$-long vector of $s_m$'s whose corresponding $u_m$'s happen to be $\bar u_j$:
\begin{equation}
\tilde s_j(u,s)=(s_m|m=1,2,...,n\mbox{ but with }u_m=\bar u_j)\in S^{n'_j(u)}.
\end{equation}
Define $V'_n\in {\cal B}(U^n\times S^n)$, so that
\begin{equation}
V'_n=\{(u,s)\in U^n\times S^n|u\in U'_n\mbox{ and }\tilde s_j(u,s)\in S'_{jn'_j(u)}\mbox{ for each }j=1,2,...,J\}.
\end{equation}

Let us follow the same logic as used from~(\ref{afact}) to~(\ref{bfact}) in the proof of (i), with appropriate substitutions, such as $J$ for $I$, $U$ for $S$, $S$ for $X$, $\nu_n$ for $\pi_n$, $\tau$ for $\sigma$, $g(\cdot,\cdot,\tau)$ for $\chi$, $g^n$ for $\chi^n$, $V'_n$ for $U'_n$,~(\ref{big12}) for~(\ref{big1}), and~(\ref{big22}) for~(\ref{big2}). We can then derive that
\begin{equation}\label{afact2}
(\nu_n\otimes g^n)(V'_n)>1-\epsilon,
\end{equation}
whereas, for any $(u,s)$ in $V'_n$,
\begin{equation}\label{bfact2}
\rho_{U\times S}(\varepsilon_{us},\tau\otimes g(\cdot,\cdot,\tau))<5\epsilon.
\end{equation}
Since~(\ref{afact2}) and~(\ref{bfact2}) are to occur at any $n$ that is large enough, we see that $\nu_n\otimes g^n$ would asymptotically resemble $(\tau\otimes g(\cdot,\cdot,\tau))^n$. Lemma~\ref{c-prob} will then lead to the asymptotic resemblance of the sequence $\nu_n\odot g^n=(\nu_n\otimes g^n)|_{S^n}$ to the sequence $(\tau\odot g(\cdot,\cdot,\tau))^n=((\tau\otimes g(\cdot,\cdot,\tau))|_S)^n$. Thus (ii) is true.

For (iii), denote the given $(s,x)$ by $u_1$. By Lemma~\ref{p-mc8}, we can make $\varepsilon_u=(u_1,\varepsilon_{u_{-1}})_n$ arbitrarily close to $\varepsilon_{u_{-1}}$ for any $u_{-1}=(u_2,u_3,...,u_n)\in U^{n-1}$ by letting $n$ be large enough. Hence, we can follow the proof of (ii) almost verbatim, with its~(\ref{new-d2}) replaced by
\begin{equation}\label{new-d3}
U'_{n-1}=\{u_{-1}\in U^{n-1}|\rho_U(\varepsilon_u,\tau)\bigvee [2\cdot \sup_{u'\in U}\max_{m=1}^n \rho_S(g(u',\varepsilon_{u_{-m}}),g(u',\tau))]<\delta\},
\end{equation}
its~(\ref{big12}) replaced by
\begin{equation}\label{big13}
(\delta_{u_1}\times \nu_{n-1})(\{u_1\}\times U'_{n-1})=\nu_{n-1}(U'_{n-1})>1-\frac{\epsilon}{2},
\end{equation}
any choice of $u\in U^n$ replaced by $u_{-1}\in U^{n-1}$, and any choice of $u\in U'_n$ replaced by $u_{-1}\in U'_{n-1}$. \qed


\noindent{\bf Proof of Theorem~\ref{T-converge}: }We prove by induction on $t'$.

First, note that $T_{[t,t-1]}\circ\sigma_t$ is merely $\sigma_t$ itself. Hence, the claim is true for $t'=t$ because by the hypothesis, we do have $\pi_{nt}$ asymptotically resembling $(T_{[t,t-1]}\circ \sigma_t)^n=\sigma_t^{\;n}$. Then, for some $t'=t,t+1,...,\bar t$, suppose the claim is true, that $\pi_{nt'}=\pi_{nt}\odot\Pi_{t''=t}^{t'-1}(\chi_{t''}^{\;n}\odot\tilde g_{t''}^{\;n})$ asymptotically resembles $\sigma_{t'}^{\;n}=(T_{[t,t'-1]}(\chi_{[t,t'-1]})\circ \sigma_t)^n$.

Assumption~\ref{g-c} on $\tilde g_{t'}(s,x,\tau)$'s equi-continuity in $\tau$ allows us to use part (ii) of Proposition~\ref{T-onestep}. By it, we would have $\pi_{nt'}\odot\chi_{t'}^{\;n}\odot\tilde g_{t'}^{\;n}$ asymptotically resembling $(\sigma_{t'}\odot \chi_{t'}\odot\tilde g_t(\cdot,\cdot,\sigma_{t'}\otimes \chi_{t'}))^n$. Since the former is merely $\pi_{n,t'+1}=\pi_{nt}\odot\Pi_{t''=t}^{t'}(\chi_{t''}^{\;n}\odot\tilde g_{t''}^{\;n})$ and the latter is $\sigma_{t'+1}^{\;\;\;n}=(T_{[tt']}(\chi_{[tt']})\circ \sigma_t)^n$, we have thus proved the claim for $t'+1$.

The induction process is now complete. \qed

\section{Proofs of Section~\ref{equilibria}}\label{app-c}

\noindent{\bf Proof of Proposition~\ref{V-convergence}: }Let us prove by induction on $t$. By~(\ref{terminal}) and~(\ref{terminal-n}), the desired result is true for $t=\bar t+1$.

At some $t=\bar t,\bar t-1,...,1$, suppose for any $\sigma_{t+1}$ and any sequence $\hat\pi_{n-1,t+1}$ that asymptotically resembles $\sigma_{t+1}^{n-1}$, the sequence $\int_{S^{n-1}}\hat\pi_{n-1,t+1}(ds_{t+1,-1})\cdot v_{n,t+1}(s_{t+1,1},\xi_{[t+1,\bar t]},,\varepsilon_{s_{t+1,-1}},\chi_{[t+1,\bar t]})$ converges to $v_{t+1}(s_{t+1,1},\xi_{[t+1,\bar t]},\sigma_{t+1},\chi_{[t+1,\bar t]})$ at a rate independent of both $s_{t+1,1}$ and $\xi_{[t+1,\bar t]}$.

Now, given the sequence $\hat\pi_{n-1,t}$ that is known to asymptotically resemble $\sigma_t^{\;n-1}$, we are to show that $\int_{S^{n-1}}\hat\pi_{n-1,t}(ds_{t,-1})\cdot v_{nt}(s_{t1},\xi_{[t\bar t]},\varepsilon_{s_{t,-1}},\chi_{[t\bar t]})$ will converge to $v_t(s_{t1},\xi_{[t\bar t]},\sigma_t,\chi_{[t\bar t]})$ at a rate independent of both $s_{t1}$ and $\xi_{[t\bar t]}$. For convenience, let $\sigma_{t+1}=T_t(\chi_t)\circ \sigma_t$.

Note that, by~(\ref{recursive}) and~(\ref{recursive-n}),
\begin{equation}\begin{array}{l}
\sup_{s_{t1}\in S,\xi_{[t\bar t]}\in ({\cal K}(S,X))^{\bar t-t+1}}\mid  v_t(s_{t1},\xi_{[t\bar t]},\sigma_t,\chi_{[t\bar t]})\\
\;\;\;\;\;\;\;\;\;\;\;\;-\int_{S^{n-1}}\hat\pi_{n-1,t}(ds_{t,-1})\cdot v_{nt}(s_{t1},\xi_{[t\bar t]},\varepsilon_{s_{t,-1}},\chi_{[t\bar t]})\mid\leq M_{n1}+M_{n2}+M_{n3},
\end{array}\end{equation}
where
\begin{equation}\label{m1d}\begin{array}{l}
M_{n1}=\sup_{(s_{t1},x_{t1})\in S\times X}\int_{S^{n-1}\times X^{n-1}}(\hat\pi_{n-1,t}\otimes \chi_t^{\;n-1})(ds_{t,-1}\times dx_{t,-1})\times\\
\;\;\;\;\;\;\;\;\;\;\;\;\;\;\;\;\;\;\;\;\;\;\;\;\times \mid \tilde f_t(s_{t1},x_{t1},\sigma_t\otimes\chi_t)-\tilde f_t(s_{t1},x_{t1},\varepsilon_{s_{t,-1}x_{t,-1}})\mid,
\end{array}\end{equation}
\begin{equation}\label{m200}\begin{array}{l}
M_{n2}=\sup_{(s_{t1},x_{t1})\in S\times X,\xi_{[t+1,\bar t]}\in ({\cal K}(S,X))^{\bar t-t}}\int_S\tilde g_t(s_{t1},x_{t1},\sigma_t\otimes\chi_t|ds_{t+1,1})\times\\
\;\;\times \mid v_{t+1}(s_{t+1,1},\xi_{[t+1,\bar t]},\sigma_{t+1},\chi_{[t+1,\bar t]})-\int_{S^{n-1}\times X^{n-1}}(\hat\pi_{n-1,t}\otimes\chi_t^{\;n-1})(ds_{t,-1}\times dx_{t,-1})\times\\
\;\;\;\;\times\Pi_{m=2}^n\int_S \tilde g_t(s_{tm},x_{tm},\varepsilon_{s_{t,-m}x_{t,-m}}|ds_{t+1,m})\cdot v_{n,t+1}(s_{t+1,1},\xi_{[t+1,\bar t]},\varepsilon_{s_{t+1,-1}},\chi_{[t+1,\bar t]})\mid,
\end{array}\end{equation}
and
\begin{equation}\label{m300}\begin{array}{l}
M_{n3}=\sup_{(s_{t1},x_{t1})\in S\times X,\xi_{[t+1,\bar t]}\in ({\cal K}(S,X))^{\bar t-t}}\int_{S^{n-1}\times X^{n-1}}(\hat\pi_{n-1,t}\otimes\chi_t^{\;n-1})(ds_{t,-1}\times dx_{t,-1})\times\\
\;\;\times  \mid [\int_S\tilde g_t(s_{t1},x_{t1},\sigma_t\otimes\chi_t|ds_{t+1,1})-\int_S\tilde g_t(s_{t1},x_{t1},\varepsilon_{s_{t,-1}x_{t,-1}}|ds_{t+1,1})]\times\\
\;\;\;\;\times\Pi_{m=2}^n\int_S\tilde g_t(s_{tm},x_{tm},\varepsilon_{s_{t,-m}x_{t,-m}}|ds_{t+1,m})\cdot  v_{n,t+1}(s_{t+1,1},\xi_{[t+1,\bar t]},\varepsilon_{s_{t+1,-1}},\chi_{[t+1,\bar t]})\mid.
\end{array}\end{equation}
We now show that each of the above three terms can be made arbitrarily small by letting $n$ be large enough.

For $M_{n1}$, define $\tilde U_{n-1}(\delta)\in {\cal B}(S^{n-1}\times X^{n-1})$ for every $\delta>0$, so that
\begin{equation}\label{an-def2}
\tilde U_{n-1}(\delta)=\{(s_{t,-1},x_{t,-1})\in S^{n-1}\times X^{n-1}|\rho_{S\times X}(\varepsilon_{s_{t,-1}x_{t,-1}},\sigma_t\otimes\chi_t)<\delta\}.
\end{equation}
From~(\ref{m1d}), we know $M_{n1}\leq M_{n11}(\delta)+M_{n12}(\delta)$ for any $\delta>0$, where
\begin{equation}
M_{n11}(\delta)=\sup_{(s_{t1},x_{t1})\in S\times X,(s_{t,-1},x_{t,-1})\in \tilde U_{n-1}(\delta)}\mid \tilde f_t(s_{t1},x_{t1},\sigma_t\otimes\chi_t)-\tilde f_t(s_{t1},x_{t1},\varepsilon_{s_{t,-1}x_{t,-1}})\mid,
\end{equation}
and
\begin{equation}\begin{array}{l}
M_{n12}(\delta)=\sup_{(s_{t1},x_{t1})\in S\times X}\int_{(S^{n-1}\times X^{n-1})\setminus\tilde U_{n-1}(\delta)} (\hat\pi_{n-1,t}\otimes \chi_t^{\;n-1})(ds_{t,-1}\times dx_{t,-1})\times\\
\;\;\;\;\;\;\;\;\;\;\;\;\;\;\;\;\;\;\;\;\;\;\;\;\times[\mid \tilde f_t(s_{t1},x_{t1},\sigma_t\otimes\chi_t)\mid+\mid\tilde f_t(s_{t1},x_{t1},\varepsilon_{s_{t,-1}x_{t,-1}})\mid].
\end{array}\end{equation}
Because Assumption~\ref{f-c} says that $\tilde f_t(s,x,\tau)$ is continuous in $\tau$ at an $(s,x)$-independent rate, we can make $M_{n11}(\delta)$ arbitrarily small by letting $\delta$ be small enough. Meanwhile, by the asymptotic resemblance of the sequence $\hat\pi_{n-1,t}$ to the sequence $\sigma_t^{\;n-1}$ and part (i) of Proposition~\ref{T-onestep}, we know that the sequence $\hat\pi_{n-1,t}\otimes\chi_t^{\;n-1}$ asymptotically resembles the sequence $(\sigma_t\otimes\chi_t)^{n-1}$. So the measure $(\hat\pi_{n-1,t}\otimes \chi_t^{\;n-1})((S^{n-1}\times X^{n-1})\setminus\tilde U_{n-1}(\delta))$ can be made arbitrarily small at any $\delta$ by letting $n$ be large enough. Since $\tilde f_t$ is bounded, this means that $M_{n12}(\delta)$ can be made arbitrarily small as well.

For $M_{n2}$, note the second integral in~(\ref{m200}) can be understood as $\hat\pi_{n-1,t+1}(s_{t1},x_{t1}|ds_{t+1,-1})=\int_{S^{n-1}}\{[(\delta_{s_{t1}x_{t1}}\times(\sigma_t^{\;n-1}\otimes \chi_t^{\;n-1}))\odot\tilde g_t^{\;n}]|_{S^{n-1}}\}(ds_{t+1,-1})$. So we have
\begin{equation}\begin{array}{l}
M_{n2}\leq \sup_{(s_{t1},x_{t1})\in S\times X,s_{t+1,1}\in S,\xi_{[t+1,\bar t]}\in ({\cal K}(S,X))^{\bar t-t}}\mid v_{t+1}(s_{t+1,1},\xi_{[t+1,\bar t]},\sigma_{t+1},\chi_{[t+1,\bar t]})\\
\;\;\;\;\;\;\;\;\;\;\;\;-\int_{S^{n-1}}\hat\pi_{n-1,t+1}(s_{t1},x_{t1}|ds_{t+1,-1})\cdot v_{n,t+1}(s_{t+1,1},\xi_{[t+1,\bar t]},\varepsilon_{s_{t+1,-1}},\chi_{[t+1,\bar t]})\mid.
\end{array}\end{equation}
Meanwhile, Assumption~\ref{g-c} allows us to use part (iii) of Proposition~\ref{T-onestep}. By the asymptotic resemblance of the sequence $\hat\pi_{n-1,t}$ to the sequence $\sigma_t^{\;n-1}$, part (iii) of Proposition~\ref{T-onestep}, and Lemma~\ref{d-prob}, we know that the sequence $\hat\pi_{n-1,t+1}(s_{t1},x_{t1})$ asymptotically resembles the sequence $\sigma_{t+1}^{n-1}$ at an $(s_{t1},x_{t1})$-independent rate. Then by the induction hypothesis where the convergence rate is also $(s_{t+1,1},\xi_{[t+1,\bar t]})$-independent, we can conclude that $M_{n2}$ can be made arbitrarily small by letting $n$ be large enough.

For $M_{n3}$, define $V_n(s_{t1},x_{t1},\varepsilon_{s_{t,-1}x_{t,-1}},\xi_{[t+1,\bar t]})$ so that
\begin{equation}\label{udef}\begin{array}{l}
V_n(s_{t1},x_{t1},\varepsilon_{s_{t,-1}x_{t,-1}},\xi_{[t+1,\bar t]})=\mid [\int_S\tilde g_t(s_{t1},x_{t1},\sigma_t\otimes\chi_t|ds_{t+1,1})\\
\;\;\;\;\;\;\;\;\;\;\;\;-\int_S\tilde g_t(s_{t1},x_{t1},\varepsilon_{s_{t,-1}x_{t,-1}}|ds_{t+1,1})]\cdot \Pi_{m=2}^n\tilde g_t(s_{tm},x_{tm},\varepsilon_{s_{t,-m}x_{t,-m}}|ds_{t+1,m})\times\\
\;\;\;\;\;\;\;\;\;\;\;\;\;\;\;\;\;\;\;\;\;\;\;\;\times v_{n,t+1}(s_{t+1,1},\xi_{[t+1,\bar t]},\varepsilon_{s_{t+1,-1}},\chi_{[t+1,\bar t]})\mid.
\end{array}\end{equation}
Then,~(\ref{m300}) can be written as
\begin{equation}\begin{array}{l}
M_{n3}=\sup_{(s_{t1},x_{t1})\in S\times X,\xi_{[t+1,\bar t]}\in ({\cal K}(S,X))^{\bar t-t}}\int_{S^{n-1}\times X^{n-1}}\times\\
\;\;\;\;\;\;\;\;\;\;\;\;\;\;\;\;\;\;\times(\hat\pi_{n-1,t}\otimes\chi_t^{\;n-1})(ds_{t,-1}\times dx_{t,-1})\cdot V_n(s_{t1},x_{t1},\varepsilon_{s_{t,-1}x_{t,-1}},\xi_{[t+1,\bar t]}).
\end{array}\end{equation}
Noting the definition of $\tilde U_{n-1}(\delta)$ in~(\ref{an-def2}) for any $\delta>0$, we see that $M_{n3}\leq M_{n31}(\delta)+M_{n32}(\delta)$, where
\begin{equation}\label{m31}\begin{array}{l}
M_{n31}(\delta)=\sup_{(s_{t1},x_{t1})\in S\times X,(s_{t,-1},x_{t,-1})\in \tilde U_{n-1}(\delta),\xi_{[t+1,\bar t]}\in ({\cal K}(S,X))^{\bar t-t}}\\
\;\;\;\;\;\;\;\;\;\;\;\;\;\;\;\;\;\;\;\;\;\;\;\;\;\;\;\;\;\;\;\;\;\;\;\;\;\;\;\;\;\;\;\;\;\;\;\; V_n(s_{t1},x_{t1},\varepsilon_{s_{t,-1}x_{t,-1}},\xi_{[t+1,\bar t]}),
\end{array}\end{equation}
and
\begin{equation}\begin{array}{l}
M_{n32}(\delta)=\sup_{(s_{t1},x_{t1})\in S\times X,\xi_{[t+1,\bar t]}\in ({\cal K}(S,X))^{\bar t-t}}\int_{(s_{t,-1},x_{t,-1})\in (S^{n-1}\times X^{n-1})\setminus\tilde U_{n-1}(\delta)}\\
\;\;\;\;\;\;\;\;\;\;\;\;\;\;\;\;\;\;\;\;\;\;\;\;\;\;\;(\hat\pi_{n-1,t}\otimes\chi_t^{\;n-1})(ds_{t,-1}\times dx_{t,-1})\cdot V_n(s_{t1},x_{t1},\varepsilon_{s_{t,-1}x_{t,-1}},\xi_{[t+1,\bar t]}).
\end{array}\end{equation}

We argue that $M_{n31}(\delta)$ can be made arbitrarily small as $\delta$ approaches $0^+$.  Due to Assumption~\ref{g-c} that $\tilde g_t(s,x,\tau)$ is continuous in $\tau$ at an $(s,x)$-independent rate, we can make $\tilde g_t(s_{t1},x_{t1},\varepsilon_{s_{t,-1}x_{t,-1}})$ for any $(s_{t,-1},x_{t,-1})\in \tilde U_{n-1}(\delta)$ arbitrarily close to $\tilde g_t(s_{t1},x_{t1},\sigma_t\otimes \chi_t)$ by rendering $\delta$ small enough, without respect to $(s_{t1},x_{t1})$. Due to its countability, we can write $S=\{\bar s_1,\bar s_2,...\}$.
Under known $s_{t1}$, $x_{t1}$, $\varepsilon_{s_{t,-1}x_{t,-1}}$, and $\xi_{[t+1,\bar t]}$, let us use the simplified notation
\begin{equation}
\gamma_i=\tilde g_t(s_{t1},x_{t1},\sigma_t\otimes \chi_t|\{\bar s_i\}),
\end{equation}
\begin{equation}
\gamma'_i=\tilde g_t(s_{t1},x_{t1},\varepsilon_{s_{t,-1}x_{t,-1}}|\{\bar s_i\}),
\end{equation}
and
\begin{equation}
v_i=\Pi_{m=2}^n\int_S \tilde g_t(s_{tm},x_{tm},\varepsilon_{s_{t,-m}x_{t,-m}}|ds_{t+1,m})\cdot v_{n,t+1}(s_{t+1,1},\xi_{[t+1,\bar t]},\varepsilon_{s_{t+1,-1}},\chi_{[t+1,\bar t]}).
\end{equation}
Then,~(\ref{udef}) can be expressed as
\begin{equation}\label{m20a}
V_n(s_{t1},x_{t1},\varepsilon_{s_{t,-1}x_{t,-1}},\xi_{[t+1,\bar t]})=\mid \sum_i \gamma_i\cdot v_i-\sum_i\gamma'_i\cdot v_i\mid.
\end{equation}

Note the $\mid v_i\mid$'s are uniformly bounded, say by $\overline v$, due to the boundedness of the $\tilde f_{t'}$'s and the finiteness of $\bar t$. Let $I$ be the set of $i$'s such that $\gamma_i\geq\gamma'_i$. Then, from~(\ref{m20a}), we have
\begin{equation}\label{newinsight}
V_n(s_{t1},x_{t1},\varepsilon_{s_{t,-1}x_{t,-1}},\xi_{[t+1,\bar t]})\leq 2\overline v\cdot \sum_{i\in I}(\gamma_i-\gamma'_i).
\end{equation}
Let $\delta$ be below $\inf_{s\neq s'}d_S(s,s')>0$. But then, $(s_{t,-1},x_{t,-1})\in \tilde U_{n-1}(\delta)$ would entail
\begin{equation}\label{consult}\begin{array}{l}
\sum_{i\in I}(\gamma_i-\gamma'_i)=\tilde g_t(s_{t1},x_{t1},\sigma_t\otimes\chi_t|\{s_i|i\in I\})-\tilde g_t(s_{t1},x_{t1},\varepsilon_{s_{t,-1}x_{t,-1}}|\{s_i|i\in I\})\\
\;\;\;\;\;\;\;\;\;\;\;\;=\tilde g_t(s_{t1},x_{t1},\sigma_t\otimes\chi_t|\{s_i|i\in I\})-\tilde g_t(s_{t1},x_{t1},\varepsilon_{s_{t,-1}x_{t,-1}}|(\{s_i|i\in I\})^\delta)<\delta.
\end{array}\end{equation}
In view of~(\ref{newinsight}), $V_n(s_{t1},x_{t1},\varepsilon_{s_{t,-1}x_{t,-1}},\xi_{[t+1,\bar t]})$ with $(s_{t,-1},x_{t,-1})\in \tilde U_{n-1}(\delta)$ can be made arbitrarily small by decreasing $\delta$ at a rate independent of $(s_{t1},x_{t1},\xi_{[t+1,\bar t]})$. In view of~(\ref{m31}), we see that $M_{n31}(\delta)$ can be made arbitrarily small by rendering $\delta$ small enough.

As noted earlier, the probability $(\hat\pi_{n-1,t}\otimes \chi_t^{\;n-1})((S^{n-1}\times X^{n-1})\setminus\tilde U_{n-1}(\delta))$ can be made arbitrarily small at any $\delta$ when $n$ is made large enough. But since $V_n(s_{t1},x_{t1},\varepsilon_{s_{t,-1}x_{t,-1}},\xi_{[t+1,\bar t]})$ is uniformly bounded, this means that $M_{n12}(\delta)$ can be made arbitrarily small as well.

Hence, all three terms can be made arbitrarily small by letting $n$ be large enough. We have thus completed the induction process.  \qed

\noindent{\bf Proof of Theorem~\ref{main}: }Given~(\ref{first}) for every $t=1,2,...,\bar t$ and $\xi_t\in{\cal K}(S,X)$, we are to verify~(\ref{second}) for every $t=1,2,...,\bar t$, $\epsilon>0$, large enough $n$, $s_{t1}\in S$, and $\xi_{[t\bar t]}\in ({\cal K}(S,X))^{\bar t-t+1}$.

First, we show that the one-time formulation of~(\ref{first}) would already imply the futility of any multi-period unilateral deviation. Another way to write the condition is, at $t'=0$, for any $t=1,2,...,\bar t-t'$ and  $\xi_{[t,t+t']}\in ({\cal K}(S,X))^{t'+1}$,
\begin{equation}\label{first1}
v_t(s_t,(\xi_{[t,t+t'-1]},\chi_{[t+t',\bar t]}),\sigma_t,\chi_{[t\bar t]})\geq v_t(s_t,(\xi_{[t,t+t']},\chi_{[t+t'+1,\bar t]}),\sigma_t,\chi_{[t\bar t]}).
\end{equation}
Now suppose~(\ref{first1}) is true for some $t'=0,1,...,\bar t-1$. We are to show its validity at $t'+1$. But by~(\ref{recursive}), for any $t=1,2,...,\bar t-t'$,
\begin{equation}\begin{array}{l}
v_t(s_t,(\xi_{[t,t+t']},\chi_{[t+t'+1,\bar t]}),\sigma_t,\chi_{[t\bar t]})-v_t(s_t,(\xi_{[t,t+t'+1]},\chi_{[t+t'+2,\bar t]}),\sigma_t,\chi_{[t\bar t]})\\
\;\;\;\;\;\;\;\;\;\;\;\;=\int_X\xi_t(s_t|dx_t)\cdot\int_S\tilde g_t(s_t,x_t,\sigma_t\otimes\chi_t|ds_{t+1})\times\\
\;\;\;\;\;\;\;\;\;\;\;\;\;\;\;\;\;\;\times [v_{t+1}(s_{t+1},(\xi_{[t+1,t+t']},\chi_{[t+t'+1,\bar t]}),T_t(\chi_t)\circ \sigma_t,\chi_{[t+1,\bar t]})\\
\;\;\;\;\;\;\;\;\;\;\;\;\;\;\;\;\;\;\;\;\;\;\;\;-v_{t+1}(s_{t+1},(\xi_{[t+1,t+t'+1]},\chi_{[t+t'+2,\bar t]}),T_t(\chi_t)\circ \sigma_t,\chi_{[t+1,\bar t]})],
\end{array}\end{equation}
which, by the induction hypothesis~(\ref{first1}), is positive. Therefore,~(\ref{first1}) is true for any $t=1,2,...,\bar t$, $t'=0,1,...,\bar t-t$, and $\xi_{[t,t+t']}\in ({\cal K}(S,X))^{t'+1}$.

Using~(\ref{first1}) multiple times, we can derive, for any $t=1,2,...,\bar t$ and $\xi_{[t\bar t]}\in ({\cal K}(S,X))^{\bar t-t+1}$,
\begin{equation}\label{first2}\begin{array}{l}
v_t(s_t,\chi_{[t\bar t]},\sigma_t,\chi_{[t\bar t]})\geq v_t(s_t,(\xi_t,\chi_{[t+1,\bar t]}),\sigma_t,\chi_{[t\bar t]})\geq v_t(s_t,(\xi_{[t,t+1]},\chi_{[t+2,\bar t]}),\sigma_t,\chi_{[t\bar t]})\\
\;\;\;\;\;\;\;\;\;\;\;\;\;\;\;\;\;\;\geq \cdots\geq v_t(s_t,(\xi_{[t,\bar t-1]},\chi_{\bar t}),\sigma_t,\chi_{[t\bar t]})\geq v_t(s_t,\xi_{[t\bar t]},\sigma_t,\chi_{[t\bar t]}).
\end{array}\end{equation}

In view of~(\ref{first2}), we would have~(\ref{second}) if for any $\epsilon$ and large enough $n$,
\begin{equation}\label{f-task}
\int_{S^{n-1}}\hat\pi_{n-1,t}(ds_{t,-1})\cdot v_{nt}(s_{t1},\chi_{[t\bar t]},\varepsilon_{s_{t,-1}},\chi_{[t\bar t]})>v_t(s_{t1},\chi_{[t\bar t]},\sigma_t,\chi_{[t\bar t]})-\frac{\epsilon}{2},
\end{equation}
and for any $\xi_{[t\bar t]}\in ({\cal K}(S,X))^{\bar t-t+1}$,
\begin{equation}\label{s-task}
\int_{S^{n-1}}\hat\pi_{n-1,t}(ds_{t,-1})\cdot v_{nt}(s_{t1},\xi_{[t\bar t]},\varepsilon_{s_{t,-1}},\chi_{[t\bar t]})<v_t(s_{t1},\xi_{[t\bar t]},\sigma_t,\chi_{[t\bar t]})+\frac{\epsilon}{2}.
\end{equation}
Both~(\ref{f-task}) and~(\ref{s-task}) would be true if
\begin{equation}\label{task}
\int_{S^{n-1}}\hat\pi_{n-1,t}(ds_{t,-1})\cdot v_{nt}(s_{t1},\xi_{[t\bar t]},\varepsilon_{s_{t,-1}},\chi_{[t\bar t]})\longrightarrow_{n\rightarrow +\infty}  v_t(s_{t1},\xi_{[t\bar t]},\sigma_t,\chi_{[t\bar t]}),
\end{equation}
at an $(s_{t1},\xi_{[t\bar t]})$-independent convergence rate. But this was provided by Proposition~\ref{V-convergence}. \qed

\section{Proofs of Section~\ref{interpretation}}\label{app-d}

\noindent{\bf Proof of Proposition~\ref{bomb1}: }Since $A$ is discrete, we can denote it by either $\{\bar a_1,\bar a_2,...\}$ or $\{\bar a_1,...,\bar a_I\}$ for some finite $I$. We work with the former only, as the latter is similarly treatable.

For any $n\in \mathbb{N}$, define \begin{equation}
N_n=\{(n_1,n_2,...)|n_i=0,1,...,n\mbox{ for each }i=1,2,...,\;\mbox{ and }\sum_{i=1}^{+\infty}n_i=n\}.
\end{equation}
For each $(n_1,n_2,...)\in N_n$, define $A^n_{n_1n_2\cdots}$ so that
\begin{equation}\label{ann-def}
A^n_{n_1n_2\cdots}=\{a\in A^n|\varepsilon_a(\{\bar a_i\})=\frac{n_i}{n}\mbox{ for any }i=1,2,...\}.
\end{equation}
Note that every $A^n_{n_1n_2\cdots}$ is symmetric, different $A^n_{n_1n_2\cdots}$'s are non-overlapping, and
\begin{equation}\label{decomp}
A^n=\bigcup_{(n_1n_2\cdots)\in N_n}A^n_{n_1n_2\cdots}.
\end{equation}
Due to the above decomposition, each $a\in A^n$ belongs to its own $A^n_{n\cdot\varepsilon_a(\{\bar a_1\}),n\cdot \varepsilon_a(\{\bar a_2\}),\cdots}$.

For any $(n_1,n_2,\cdots)\in N_n$, the set $A^n_{n_1n_2\cdots}$ contains $n!/(\prod_{i=1}^{+\infty}n_i!)$ distinct members of $A^n$, say $a_1,...,a_{n!/(\prod_{i=1}^{+\infty}n_i!)}$. In addition, every $a_k$ is of the form $\psi a_1$ for some $\psi\in\Psi_n$. Thus, due to $q_n$'s symmetry, for $k=1,2,...,n!/(\prod_{i=1}^{+\infty}n_i!)$,
\begin{equation}\label{ratio}
q_n(\{a_k\})=\frac{\prod_{i=1}^{+\infty}n_i!}{n!}\cdot q_n(A^n_{n_1n_2\cdots}).
\end{equation}
Suppose $n_i\geq 1$ for some $i=1,2,...$. Then, exactly $(n-1)!/((n_i-1)!\cdot\prod_{j\neq i}n_j!)$ of the $a_k$'s will have $a_{k1}=\bar a_i$. Therefore, for any such $a_k$,
\begin{equation}\label{observe}
q_n((\{\bar a_i\}\times A^{n-1})\cap A^n_{n_1n_2\cdots})=\frac{(n_i-1)!\cdot\prod_{j\neq i}n_j!}{(n-1)!}\cdot q_n(\{a_k\})=\frac{n_i}{n}\cdot q_n(A^n_{n_1n_2\cdots}),
\end{equation}
where the second equality stems from~(\ref{ratio}). The above left- and right-hand sides are certainly equated as well when $n_i=0$. Combine~(\ref{decomp}) and~(\ref{observe}), and we can obtain
\begin{equation}\label{decomp-q}
q_n(\{\bar a_i\}\times A^{n-1})=\sum_{(n_1,n_2,...)\in N_n}\frac{n_i}{n}\cdot q_n(A^n_{n_1n_2\cdots}).
\end{equation}

On the other hand, we have $\min_{i\neq j}d_A(\bar a_i,\bar a_j)>0$ due to $A$'s discreteness. Suppose $\epsilon>0$ is small enough to be strictly below this constant. Then by the nature of the Prohorov metric,
$a\in A^n$ would satisfy $\rho_A(\varepsilon_a,p)<\epsilon$ if and only if
\begin{equation}\label{revelation}
\sum_{i=1}^{+\infty}\mid \varepsilon_a(\{\bar a_i\})-p(\{\bar a_i\})\mid<2\epsilon,
\end{equation}
and hence only if
\begin{equation}\label{revelation-b}
\max_{i=1}^{+\infty}\mid \varepsilon_a(\{\bar a_i\})-p(\{\bar a_i\})\mid<\epsilon.
\end{equation}
Since the sequence $q_n$ asymptotically resembles $p$, for any $\epsilon>0$ that is strictly below $\min_{i\neq j}d_A(\bar a_i,\bar a_j)>0$, we can pick $n$ large enough so that~(\ref{a78a}) and~(\ref{b78a}) in the proof of Lemma~\ref{d-prob} are true. Define $N'_n\subseteq N_n$ so that for any $(n_1,n_2,...)\in N'_n$,
\begin{equation}
\sum_{i=1}^{+\infty}\mid \frac{n_i}{n}-p(\{\bar a_i\})\mid<2\epsilon,\;\mbox{ and hence }\;\max_{i=1}^{+\infty}\mid \frac{n_i}{n}-p(\{\bar a_i\})\mid<\epsilon.
\end{equation}
Due to~(\ref{ann-def}),~(\ref{decomp}), and~(\ref{revelation}), we have the following for the $A'_n$ defined in~(\ref{b78a}):
\begin{equation}\label{revelation-a}
A'_n=\bigcup_{(n_1,n_2,...)\in N'_n}A^n_{n_1n_2\cdots}.
\end{equation}

Now for any $i=1,2,...$, we have
\begin{equation}\begin{array}{l}
\mid q_n|_A(\{\bar a_i\})-p(\{\bar a_i\})\mid=\mid q_n(\{\bar a_i\}\times A^{n-1})-p(\{\bar a_i\})\mid\\
\;\;\;\;\;\;\;\;\;\;\;\;=\mid (\sum_{(n_1,n_2,...)\in N'_n}+\sum_{(n_1,n_2,...)\in N_n\setminus N'_n})n_i\cdot q_n(A^n_{n_1n_2\cdots})/n-p(\{\bar a_i\})\mid\\
\;\;\;\;\;\;\;\;\;\;\;\;\leq q_n(A'_n)\cdot\mid n_i/n-p(\{\bar a_i\})\mid+q_n(A^n\setminus A'_n)<2\epsilon.
\end{array}\end{equation}
Here, the first equality comes from the definition of marginal probability, the second equality comes from~(\ref{decomp-q}), the first inequality can be attributed to~(\ref{revelation-a}), and the last inequality is due to~(\ref{a78a}) and~(\ref{revelation-b}). Thus, for every $a\in A$, we have $\lim_{n\rightarrow +\infty}q_n|_A(\{a\})=p(\{a\})$. \qed

\noindent{\bf Proof Proposition~\ref{bomb2}: }For the time being, it does not matter whether $A=\{\bar a_1,\bar a_2,...\}$ or $\{\bar a_1,...,\bar a_I\}$ for some finite $I$. The first few steps are the same as those in the proof of Lemma~\ref{d-prob}. For any $\epsilon>0$ and $n$ large enough, we can have~(\ref{a78a}) to~(\ref{d78a}) as in that proof.

Fix $i=1,2,...$ with $p(\{\bar a_i\})>0$. Due to~(\ref{c78a}),
\begin{equation}
A'_n\cap (\{\bar a_i\}\times A^{n-1})\subseteq (A\times A''_{n-1})\cap (\{\bar a_i\}\times A^{n-1})=\{\bar a_i\}\times A''_{n-1},
\end{equation}
where $A'_n$ is defined in~(\ref{b78a}) and $A''_{n-1}$ is defined in~(\ref{d78a}). Thus,
\begin{equation}\label{lucky1}\begin{array}{l}
q_n(\{\bar a_i\}\times A''_{n-1})\geq q_n(A'_n\cap (\{\bar a_i\}\times A^{n-1}))=q_n((\{\bar a_i\}\times A^{n-1})\setminus (A^n\setminus A'_n))\\
\;\;\;\;\;\;\;\;\;\;\;\;\geq q_n(\{\bar a_i\}\times A^{n-1})-q_n(A^n\setminus A'_n)> q_n(\{\bar a_i\}\times A^{n-1})-\epsilon,
\end{array}\end{equation}
where the last inequality is due to~(\ref{a78a}).

Since $q_n$ is symmetric, we know from Proposition~\ref{bomb1} that, when $n$ is large enough,
\begin{equation}\label{lucky2}
q_n(\{\bar a_i\}\times A^{n-1})=q_n|_A(\{\bar a_i\})>\frac{p(\{\bar a_i\})}{2}>0.
\end{equation}
Combining~(\ref{lucky1}) and~(\ref{lucky2}), we can obtain
\begin{equation}\label{okla}
q_{n,A}|_{A^{n-1}}(\bar a_i|A''_{n-1})=\frac{q_n(\{\bar a_i\}\times A''_{n-1})}{q_n(\{\bar a_i\}\times A^{n-1})}>1-\frac{\epsilon}{q_n(\{\bar a_i\}\times A^{n-1})}>1-\frac{2\epsilon}{p(\{\bar a_i\})}.
\end{equation}
With $A''_{n-1}$'s definition in~(\ref{d78a}), we get $q_{n,A}|_{A^{n-1}}(\bar a_i|\cdot)$'s asymptotic resemblance to $p^{n-1}$. \qed

\section{Developments in Section~\ref{stationary}}\label{app-e}

\noindent{\bf Proof of Theorem~\ref{main-s}: }Let $\epsilon>0$ be fixed. Given $t=1,2,...$ and $\chi\in {\cal K}(S,X)$, we use $\chi^t$ to denote $(\chi,\chi,...,\chi)\in ({\cal K}(S,X))^t$.
From~(\ref{bounded-ss}), we know
\begin{equation}
\mid v^\infty_n(s_1,\xi_{[1\infty]},\varepsilon_{s_{-1}},\chi^\infty)-v^t_n(s_1,\xi_{[1t]},\varepsilon_{s_{-1}},\chi^t)\mid\leq \frac{\bar\alpha^t\cdot\bar f}{1-\bar\alpha}.
\end{equation}
Hence, when $t\geq \ln(6\bar f/(\epsilon\cdot(1-\bar\alpha)))/\ln(1/\bar\alpha)+1$,
\begin{equation}
v^\infty_n(s_1,\chi^\infty,\varepsilon_{s_{-1}},\chi^\infty)>v^t_n(s_1,\chi^t,\varepsilon_{s_{-1}},\chi^t)-\frac{\epsilon}{6},
\end{equation}
and
\begin{equation}
v^\infty_n(s_1,\xi_{[1\infty]},\varepsilon_{s_{-1}},\chi^\infty)<v^t_n(s_1,\xi_{[1t]},\varepsilon_{s_{-1}},\chi^t)+\frac{\epsilon}{6},
\end{equation}
for every $s_1\in S$, $s_{-1}\in S^{n-1}$, and $\xi_{[1\infty]}\in ({\cal K}(S,X))^\infty$. Therefore, we need merely to select such a large $t$ and show that, when $n$ is large enough,
\begin{equation}\label{merely}
\int_{S^{n-1}}\hat\pi_{n-1}(ds_{-1})\cdot v^t_n(s_1,\chi^t,\varepsilon_{s_{-1}},\chi^t)\geq \int_{S^{n-1}}\hat\pi_{n-1}(ds_{-1})\cdot  v^t_n(s_1,\xi_{[1t]},\varepsilon_{s_{-1}},\chi^t)-\frac{2\epsilon}{3},
\end{equation}
for every $s_1\in S$ and $\xi_{[1t]}\in ({\cal K}(S,X))^t$.

Since $(\chi,\sigma)$ poses as an equilibrium for $\Gamma$, we know~(\ref{first-s}) is true. Another way to write the condition is, at $t'=0$, for any $\xi_{[1,t'+1]}\in ({\cal K}(S,X))^{t'+1}$,
\begin{equation}\label{first1-s}
v^\infty(s,(\xi_{[1t']},\chi^\infty),\sigma,\chi^\infty)\geq v^\infty(s,(\xi_{[1,t'+1]},\chi^\infty),\sigma,\chi^\infty).
\end{equation}
Now suppose~(\ref{first1-s}) is true for some $t'=0,1,...$. We are to show its validity at $t'+1$. By~(\ref{recursive-s}),~(\ref{compatible-s}), and the uniform convergence of $v^t(s,\xi_{[1t]},\sigma,\chi^t)$ to $v^\infty(s,\xi_{[1\infty]},\sigma,\chi^\infty)$, we have
\begin{equation}\label{recursive-s-inf}\begin{array}{l}
v^\infty(s,\xi_{[1\infty]},\sigma,\chi^\infty)=\int_X \xi_1(s|dx)\cdot[\tilde f(s,x,\sigma\otimes \chi)\\
\;\;\;\;\;\;\;\;\;\;\;\;+\bar\alpha\cdot\int_S \tilde g(s,x,\sigma\otimes \chi|ds')\cdot v^\infty(s',\xi_{[2\infty]},\sigma,\chi^\infty)].
\end{array}\end{equation}
Therefore,
\begin{equation}\begin{array}{l}
v^\infty(s,(\xi_{[1,t'+1]},\chi^\infty),\sigma,\chi^\infty)-v^\infty(s,(\xi_{[1,t'+2]},\chi^\infty),\sigma,\chi^\infty)\\
\;\;\;\;\;\;\;\;\;\;\;\;=\int_X\xi_1(s|dx)\cdot\int_S\tilde g(s,x,\sigma\otimes\chi|ds')\times \\ \;\;\;\;\;\;\;\;\;\;\;\;\;\;\;\;\;\;\times [v^\infty(s',(\xi_{[2,t'+1]},\chi^\infty),\sigma,\chi^\infty)-v^\infty(s',(\xi_{[2,t'+2]},\chi^\infty),\sigma,\chi^\infty)],
\end{array}\end{equation}
which, by the induction hypothesis~(\ref{first1-s}), is positive. Therefore,~(\ref{first1-s}) is true for $t'=0,1,...$.

By using~(\ref{first1-s}) multiple times, we can derive that, for any $\xi_{[1t]}\in ({\cal K}(S,X))^t$,
\begin{equation}\label{first2-s}\begin{array}{l}
v^\infty(s,\chi^\infty,\sigma,\chi^\infty)\geq v^\infty(s,(\xi_1,\chi^\infty),\sigma,\chi^\infty)\geq v^\infty(s,(\xi_{[12]},\chi^\infty),\sigma,\chi^\infty)\\
\;\;\;\;\;\;\;\;\;\;\;\;\;\;\;\;\;\;\geq \cdots\geq v^\infty(s,(\xi_{[1,t-1]},\chi^\infty),\sigma_t,\chi^\infty)\geq v^\infty(s,(\xi_{[1t]},\chi^\infty),\sigma,\chi^\infty).
\end{array}\end{equation}
Also, we know from~(\ref{bounded-s}) that
\begin{equation}\label{bounded2-s}
\mid v^\infty(s,\zeta_{[1\infty]},\sigma,\chi^\infty)-v^t(s,\zeta_{[1t]},\sigma,\chi^t)\mid\leq \frac{\bar\alpha^t\cdot\bar f}{1-\bar\alpha},
\end{equation}
regardless of the $\zeta_{[1\infty]}\in ({\cal K}(S,X))^\infty$ chosen. However,~(\ref{first2-s}) and~(\ref{bounded2-s}) would together lead to
\begin{equation}\label{first-s-tau}
v^t(s,\chi^t,\sigma,\chi^t)-v^t(s,\xi_{[1t]},\sigma,\chi^t)\geq -\frac{2\bar\alpha^{t-1}\cdot\bar f}{1-\bar\alpha}\geq -\frac{\epsilon}{3},
\end{equation}
for any $s\in S$ and $\xi_{[1t]}\in ({\cal K}(S,X))^t$.

In the presence of Assumptions~\ref{g-c} and~\ref{f-c} for the corresponding $t$-period games, Proposition~\ref{V-convergence} applies. Plus, it has been hypothesized that the sequence $\hat\pi_{n-1}$ asymptotically resembles the sequence $\sigma^{n-1}$. Therefore, for $n$ large enough,
\begin{equation}\label{m1}
\int_{S^{n-1}}\hat\pi_{n-1}(ds_{-1})\cdot v^t_n(s_1,\chi^t,\varepsilon_{s_{-1}},\chi^t)>v^t(s_1,\chi^t,\sigma,\chi^t)-\frac{\epsilon}{6},
\end{equation}
regardless of the choice on $s_1\in S$, and
\begin{equation}\label{s1}
\int_{S^{n-1}}\hat\pi_{n-1}(ds_{-1})\cdot v^t_n(s_1,\xi_{[1t]},\varepsilon_{s_{-1}},\chi^t)<v^t(s_1,\xi_{[1t]},\sigma,\chi^t)+\frac{\epsilon}{6},
\end{equation}
regardless of the choices on $s_1\in S$ and $\xi_{[1t]}\in ({\cal K}(S,X))^t$. Put~(\ref{first-s-tau}) to~(\ref{s1}) together, and we would obtain~(\ref{merely}). \qed

For something akin to the second example in Section~\ref{interpretation}, we need to consider the following invariant equation involving $\pi_n\in {\cal P}(S^n)$, which is inspirable from its finite-$t$ version~(\ref{second-e}):
\begin{equation}\label{second-es}
\pi_n=\pi_n\odot \chi^n\odot \tilde g^n.
\end{equation}
Suppose~(\ref{second-es}) has a solution that asymptotically resembles $\sigma^n$, then we can let $\hat\pi_{n-1}=\pi_n|_{S^{n-1}}$. By Lemma~\ref{d-prob}, this choice would satisfy the condition in Theorem~\ref{main-s}. Its meaning is also clear---let players update their estimates on other players' states most precisely without using their own state information.

When the state space $S$ is finite, we again have an extended version much like Theorem~\ref{main-f}. If we succeed in finding a satisfactory $\pi_n$, we would be able to make the third choice of letting each $\hat\pi_{n-1}(s_1|\cdot)$ in the extended version be the conditional probability $\pi_{n,S}|_{S^{n-1}}(s_1|\cdot)$. Propositions~\ref{bomb1} and~\ref{bomb2} would then lead to the satisfaction of the corresponding condition in the extended version. The third choice here again means that players update other players' states in the most accurate Bayesian fashion.

The above second and third choices are premised on the following conjecture.

\begin{conjecture}\label{conj1}
	Suppose $\chi\in {\cal K}(S,X)$, $\tilde g\in {\cal G}(S,X)$ enjoys the continuity of $\tilde g(s,x,\tau)$ in $\tau$ at an $(s,x)$-independent rate, and $\sigma\in{\cal P}(S)$ is an solution to the invariant equation $\sigma
	=\sigma\odot\chi\odot \tilde g(\cdot,\cdot,\sigma\otimes\chi)$ as defined by~(\ref{T-def-s}) and~(\ref{compatible-s}). Then, there would exist a sequence $\pi_n$ so that for each $n\in \mathbb{N}$, $\pi_n$ as a member of ${\cal P}(S^n)$ satisfies the invariant equation $\pi_n=\pi_n\odot \chi^n\odot \tilde g^n$ as indicated by~(\ref{second-es}), and yet the sequence asymptotically resembles the sequence $\sigma^n$.
\end{conjecture}

To tackle this conjecture, one may be tempted to show that (i) iteratively applying $\sigma_{t+1}=\sigma_t\odot \chi\odot\tilde g(\cdot,\cdot,\sigma_t\otimes\chi)$ leads to the convergence of $\sigma_t$ to an invariant $\sigma$, (ii) iteratively applying $\pi_{n,t+1}=\pi_{nt}\odot \chi^n\odot\tilde g^n$ leads to the convergence of $\pi_{nt}$ to an invariant $\pi_n$ for each $n$, and (iii) these convergence results along with the asymptotic resemblance of each $\pi_{nt}$ to $\sigma_t^{\;n}$ would lead to that of $\pi_n$ to $\sigma^n$. So far, (i) and (ii) still elude us. On the other hand, something slightly weaker than (iii) can be achieved.

\begin{proposition}\label{bomb3}
	Let $A$ be a separable metric space, and $p_i$ for $i\in \mathbb{N}$ and $p$ be members of ${\cal P}(A)$. Also, for each $n\in \mathbb{N}$, let $q_{ni}$ for $i\in\mathbb{N}$ and $q_n$ be members of ${\cal P}(A^n)$. Suppose $p_i$ converges to $p$, $q_{ni}$ converges to $q_n$ for each $n\in\mathbb{N}$, and $q_{ni}$ asymptotically resembles $p_i^{\;n}$.
	Then, in either situation (a) where the convergence of $q_{ni}$ to $q_n$ is at an $n$-independent rate or situation (b) where the asymptotic resemblance of $q_{ni}$ to $p_i^{\;n}$ is at an $i$-independent rate, the sequence $q_n$ would asymptotically resemble the sequence $p^n$.
\end{proposition}
\noindent{\bf Proof of Proposition~\ref{bomb3}: }Let $\epsilon>0$ be given. Since $p_i$ converges to $p$, we have
\begin{equation}\label{best1}
\rho_A(p,p_i)<\frac{\epsilon}{2},
\end{equation}
as long as $i$ is large enough.

Suppose situation (a) is true. By the equi-$n$ convergence of $q_{ni}$ to $q_n$, we can pick $i$ large enough to ensure both~(\ref{best1}) and for any $n\in \mathbb{N}$,
\begin{equation}\label{best2}
q_n((A'_n)^{\epsilon/4})>q_{ni}(A')-\frac{\epsilon}{2},\hspace*{.5in}\forall A'_n\in {\cal B}(A^n).
\end{equation}
At such a fixed $i\in\mathbb{N}$, due to the asymptotic resemblance of $q_{ni}$ to $p_i^{\;n}$, we can let $n$ be large enough so that
\begin{equation}\label{best3}
q_{ni}(\{a\in A^n|\rho_A(\varepsilon_a,p_i)<\frac{\epsilon}{4}\})>1-\frac{\epsilon}{2}.
\end{equation}
Suppose situation (b) is true. Due to the equi-$i$ asymptotic resemblance of $q_{ni}$ to $p_i^{\;n}$, we can pick $n$ large enough to ensure~(\ref{best3}) for any $i\in\mathbb{N}$. By the convergence of $q_{ni}$ to $q_n$, we can then pick $i$ large enough to ensure~(\ref{best1}), as well as~(\ref{best2}) for the current $n\in\mathbb{N}$.

Either way, without loss of generality, we can suppose $d_{A^n}(a,a')\geq \max_{m=1}^n d_A(a_m,a'_m)$. Then, due to Lemma~\ref{newlemma},
\begin{equation}\label{best4}
(\{a\in A^n|\rho_A(\varepsilon_a,p_i)<\frac{\epsilon}{4}\})^{\epsilon/4}\subseteq \{a\in A^n|\rho_A(\varepsilon_a,p_i)<\frac{\epsilon}{2}\}.
\end{equation}
Now we can deduce that
\begin{equation}\begin{array}{l}
q_n(\{a\in A^n|\rho_A(\varepsilon_a,p)<\epsilon\})>q_n(\{a\in A^n|\rho_A(\varepsilon_a,p_i)<\epsilon/2\})\\
\;\;\;\;\;\;\;\;\;\;\;\;>q_n((\{a\in A^n|\rho_A(\varepsilon_a,p_i)<\epsilon/4\})^{\epsilon/4})\\
\;\;\;\;\;\;\;\;\;\;\;\;>q_{ni}(\{a\in A^n|\rho_A(\varepsilon_a,p_i)<\epsilon/4\})-\epsilon/2>1-\epsilon,
\end{array}\end{equation}
where the first inequality is due to~(\ref{best1}), the second inequality is due to~(\ref{best4}), the third inequality is due to~(\ref{best2}), and the last inequality is due to~(\ref{best3}). Therefore, the sequence $q_n$ asymptotically resembles the sequence $p^n$.\qed

Like Propositions~\ref{bomb1} and~\ref{bomb2}, Proposition~\ref{bomb3} also helps to bolster the legitimacy of the asymptotic resemblance concept. 

\section{Developments in Section~\ref{existence}}\label{app-f}

\subsection{The Transient Case}\label{app-f-t}

By the discreteness of $S$, every $\chi_t(s|X')$ is automatically continuous and hence measurable in $s$, and hence ${\cal K}(S,X)$ is not only a member of $({\cal P}(X))^S$, but also the latter itself.
Denote the space $({\cal P}(S))^{\bar t-1}$ by ${\cal S}$ and the space $(({\cal P}(X))^S)^{\bar t}=({\cal K}(S,X))^{\bar t}$ by ${\cal X}$. Let ${\cal U}={\cal S}\times {\cal X}$. Define a correspondence $H:{\cal U}\Rightarrow {\cal U}$, so that for any $\sigma_{[2\bar t]}\in {\cal S}$ and $\chi_{[1\bar t]}\in {\cal X}$,
\begin{equation}
H(\sigma_{[2\bar t]},\chi_{[1\bar t]})=H^S(\sigma_{[2\bar t]},\chi_{[1\bar t]})\times H^X(\sigma_{[2\bar t]},\chi_{[1\bar t]}),
\end{equation}
where
\begin{equation}\label{hs-def}
H^S(\sigma_{[2\bar t]},\chi_{[1\bar t]})=\{\sigma'_{[2\bar t]}\in {\cal S}|\sigma'_t=T_{t-1}(\chi_{t-1})\circ \sigma_{t-1},\;\forall t=2,3,...,\bar t\},
\end{equation}
and
\begin{equation}\label{hx-def}
H^X(\sigma_{[2\bar t]},\chi_{[1\bar t]})=\{\chi'_{[1\bar t]}\in {\cal X}|\chi'_t(s_t|\tilde X_t(s_t,\sigma_t,\chi_{[t\bar t]}))=1,\;\forall t=1,2,...,\bar t,s_t\in S\}.
\end{equation}

A fixed point $(\sigma_{[2\bar t]},\chi_{[1\bar t]})$ for $H$ would provide a Markov equilibrium $\chi_{[1\bar t]}$ for $\Gamma(\sigma_1)$ in the sense of~(\ref{first-alt}), with $\sigma_{[2\bar t]}$ supplying the deterministic pre-action environment pathway from period 2 to $\bar t$ that is generated from all players adopting policy $\chi_{[1\bar t]}$.
We are to use Kakutani-Fan-Glicksberg fixed point theorem to prove the existence of a fixed point for $H$. But first let us work out a couple of useful continuity results.

\begin{proposition}\label{C-onestep}
	\indent\M (i) $\sigma\otimes\chi$ is continuous in both $\sigma\in {\cal P}(S)$ and $\chi\in ({\cal P}(X))^S$.\\
	When $g\in {\cal G}(S,X)$ satisfies that $g(s,x,\tau)$ is continuous in $\tau$ at an $(s,x)$-independent rate,\\
	\indent\M (ii) $\sigma\odot \chi\odot g(\cdot,\cdot,\sigma\otimes\chi)$ is continuous in both $\sigma\in {\cal P}(S)$ and $\chi\in ({\cal P}(X))^S$.
\end{proposition}
\noindent{\bf Proof of Proposition~\ref{C-onestep}: }We first prove (i) by showing that, for any two sequences $\sigma_m$ and $\chi_m$ that converge to $\sigma$ and $\chi$, respectively, the sequence $\sigma_m\otimes\chi_m$ would converge to $\sigma\otimes\chi$. In the following, we omit detailed reasonings behind some of the steps, as they have appeared in the proof of Proposition~\ref{T-onestep}.

Fix some $\epsilon\in (0,1)$. We can identify some $I$ of its points $\bar s_1,\bar s_2,...,\bar s_I$, so that~(\ref{usedlater}) is true. For convenience, let $\bar S'=\{\bar s_1,\bar s_2,...,\bar s_I\}$ and $\bar S''=S\setminus\bar S'$. It is known that the distance $d_S(\bar S',\bar S'')=\inf_{s'\in \bar S',s''\in \bar S''}d_S(s',s'')>0$. For $i,j=1,2,...,I$, use $d_{ij}$ for $d_S(\bar s_i,\bar s_j)$ and $\sigma_i$ for $\sigma(\{\bar s_i\})$. Again, define $\delta$ through~(\ref{delta-def}), whose strict positivity is guaranteed.

As $\sigma_m$ and $\chi_m$ converge to $\sigma$ and $\chi$, respectively, for large enough $m$, we have
\begin{equation}\label{aa1}
\rho_S(\sigma,\sigma_m)<\delta,
\end{equation}
and
\begin{equation}\label{bb1}
\rho_X(\chi(\bar s_i),\chi_m(\bar s_i))<\delta,\hspace*{.5in}\forall i=1,2,...,I.
\end{equation}
Together with the fact that $\delta\leq d_S(\bar S',\bar S'')\wedge(\min_{i\neq j}d_{ij})$,~(\ref{aa1}) would result with
\begin{equation}\label{later-d0}
\sigma_i-\delta<\sigma_m(\{\bar s_i\})<\sigma_i+\delta.
\end{equation}
Meanwhile,~(\ref{bb1}) would lead to
\begin{equation}\label{big30}
\chi_m(\bar s_i|X')<\chi(\bar s_i|(X')^\delta)+\delta,\hspace*{.5in}\forall i=1,2,...,I.
\end{equation}

Any $U'\in {\cal B}(S\times X)$ still enjoys the decomposition provided in~(\ref{aaa}), that $U'=(\bigcup_{i=1}^I\{\bar s_i\}\times X'_i)\bigcup U''$, where $X'_i\in {\cal B}(X)$ for $i=1,2,...,I$, while $U''$ is such that $s''\in \bar S''$ for any $(s'',x'')\in U''$. This would result in the same~(\ref{a2a}). On the other hand, from the right half of~(\ref{later-d0}) and~(\ref{big30}),
\begin{equation}\label{aaa0}\begin{array}{l}
(\sigma_m\otimes \chi_m)(\{\bar s_i\}\times X'_i)=\sigma_m(\{\bar s_i\})\cdot \chi_m(\bar s_i|X'_i)<(\sigma_i+\delta)\cdot [\chi(\bar s_i|(X'_i)^\delta)+\delta]\\
\;\;\;\;\;\;\leq(\sigma\otimes\chi)(\{\bar s_i\}\times (X'_i)^\delta)+2\delta+\delta^2<(\sigma\otimes\chi)(\{\bar s_i\}\times (X'_i)^\delta)+3\delta,
\end{array}\end{equation}
where the last inequality is due to our choice that $\delta\leq\epsilon/I<1$. Meanwhile,
\begin{equation}\label{bbb0}\begin{array}{l}
(\sigma_m\otimes\chi_m)(U'')\leq (\sigma_m\otimes\chi_m)(\bar S''\times X)=\sigma_m(\bar S'')=1-\sum_{i=1}^I \sigma_m(\{\bar s_i\})\\
\;\;\;\;\;\;\;\;\;\;\;\;\;\;\;\;\;\;<1-\sum_{i=1}^I\sigma_i+I\delta<\epsilon+I\delta,
\end{array}\end{equation}
where the second-to-last inequality is due to the left half of~(\ref{later-d0}) and the last one is due to~(\ref{usedlater}). By combining~(\ref{aaa}),~(\ref{a2a}),~(\ref{aaa0}), and~(\ref{bbb0}), we can obtain
\begin{equation}
(\sigma_m\otimes\chi_m)(U')<(\sigma\otimes\chi)((U')^\delta)+\epsilon+4I\delta.
\end{equation}
Thus,
\begin{equation}\label{bfact0}
\rho_{S\times X}(\sigma_m\otimes\chi_m,\sigma\otimes\chi)<\epsilon+4I\delta\leq 5\epsilon.
\end{equation}
Since~(\ref{bfact0}) is to occur at any $m$ that is large enough, we see that (i) is true.

We then prove (ii). Again, suppose two sequences $\sigma_m$ and $\chi_m$ converge to $\sigma$ and $\chi$, respectively. From (i), we know $\sigma_m\otimes \chi_m$ converges to $\sigma\otimes\chi$ too. According to (87) of Yang \cite{Y11}, for any $m$,
\begin{equation}
\rho_X(\sigma_m\odot\chi_m,\sigma\odot\chi)=\rho_X((\sigma_m\otimes\chi_m)|_X,(\sigma\otimes\chi)|_X)\leq \rho_{S\times X}(\sigma_m\otimes\chi_m,\sigma\otimes\chi).
\end{equation}
Hence, there is also the convergence of $\sigma_m\odot \chi_m$ to $\sigma\odot\chi$.

On the other hand, the discrete property of $S\times X$ means $g(\cdot,\cdot,\tau)$ is a member of $({\cal P}(S))^{S\times X}$ for any fixed $\tau\in {\cal P}(S\times X)$. Now (i) and the fact that $g(s,x,\tau)$ is continuous in $\tau$ at an $(s,x)$-independent rate would together mean that, the sequence $g(\cdot,\cdot,\sigma_m\otimes\chi_m)$ in $({\cal P}(S))^{S\times X}$ converges to $g(\cdot,\cdot,\sigma\otimes\chi)$.

Let us use the convergence of $\sigma_m\odot \chi_m$ to $\sigma\odot\chi$ under proper substitutions. As $S\times X$ has been noted to be discrete, we can treat it as $S$ in the convergence result. Also, let us treat $\sigma_m\otimes \chi_m$ as $\sigma_m$, $\sigma\otimes\chi$ as $\sigma$, $S$ as $X$, $g(\cdot,\cdot,\sigma_m\otimes\chi_m)$ as $\chi_m$, and $g(\cdot,\cdot,\sigma\otimes\chi)$ as $\chi$.

From (i) on the convergence of $\sigma_m\otimes \chi_m$ to $\sigma\otimes\chi$, now viewed as that of $\sigma_m$ to $\sigma$, as well as the convergence of $g(\cdot,\cdot,\sigma_m\otimes\chi_m)$ to $g(\cdot,\cdot,\sigma\otimes\chi)$, now viewed as that of $\chi_m$ to $\chi$, we can conclude that $(\sigma_m\otimes\chi_m)\odot g(\cdot,\cdot,\sigma_m\otimes\chi_m)=\sigma_m\odot\chi_m\odot g(\cdot,\cdot,\sigma_m\otimes\chi_m)$ would converge to $(\sigma\otimes\chi)\odot g(\cdot,\cdot,\sigma\otimes\chi)=\sigma\odot\chi\odot g(\cdot,\cdot,\sigma\otimes\chi)$. Thus, (ii) is true as well.\qed

\begin{proposition}\label{V-continuity}
	For each $t=1,2,...,\bar t+1$, the value $v_t(s_t,\xi_{[t\bar t]},\sigma_t,\chi_{[t\bar t]})$ defined in~(\ref{recursive}) is continuous in $\sigma_t\in {\cal S}$ and $\chi_{[t\bar t]}\in {\cal X}$ at an $(s_t,\xi_{[t\bar t]})$-independent rate.
\end{proposition}
\noindent{\bf Proof of Proposition~\ref{V-continuity}: }We use induction on $t$. By~(\ref{terminal}), our claim is certainly true for $t=\bar t+1$. Suppose for some $t=\bar t,\bar t-1,...,1$, the function $v_{t+1}(s_{t+1},\xi_{[t+1,\bar t]},\sigma_{t+1},\chi_{[t+1,\bar t]})$ is continuous in $\sigma_{t+1}$ and $\chi_{[t+1,\bar t]}$ at a rate independent of $s_{t+1}$ and $\xi_{[t+1,\bar t]}$.

Now we prove the continuity in $\sigma_t$ and $\chi_{[t\bar t]}$ at time $t$. From~(\ref{recursive}), we have
\begin{equation}
\sup_{s_t\in S,\xi_{[t\bar t]}\in (({\cal P}(X))^S)^{\bar t-t+1}}\mid v_t(s_t,\xi_{[t\bar t]},\sigma_t,\chi_{[t\bar t]})-v_t(s_t,\xi_{[t\bar t]},\sigma'_t,\chi'_{[t\bar t]})\mid\leq M_1+M_2+M_3,
\end{equation}
where
\begin{equation}
M_1=\sup_{(s_t,x_t)\in S\times X}\mid \tilde f_t(s_t,x_t,\sigma_t\otimes \chi_t)-\tilde f_t(s_t,x_t,\sigma'_t\otimes \chi'_t)\mid,
\end{equation}
\begin{equation}\label{m22}\begin{array}{l}
M_2=\sup_{(s_t,x_t)\in S\times X,\;\xi_{[t+1,\bar t]}\in(({\cal P}(X))^S)^{\bar t-t} }\mid [\int_S\tilde g_t(s_t,x_t,\sigma_t\otimes \chi_t|ds_{t+1})\\
\;\;\;\;\;\;\;\;\;\;\;\;-\int_S\tilde g_t(s_t,x_t,\sigma'_t\otimes \chi'_t|ds_{t+1})]\cdot v_{t+1}(s_{t+1},\xi_{[t+1,\bar t]},T_t(\chi_t)\circ\sigma_t,\chi_{[t+1,\bar t]})\mid,
\end{array}\end{equation}
and
\begin{equation}\label{m33}\begin{array}{l}
M_3=\sup_{(s_t,x_t)\in S\times X,\;\xi_{[t+1,\bar t]}\in(({\cal P}(X))^S)^{\bar t-t}}\int_S\tilde g_t(s_t,x_t,\sigma'_t\otimes \chi'_t|ds_{t+1})\times\\
\;\;\;\times \mid v_{t+1}(s_{t+1},\xi_{[t+1,\bar t]},T_t(\chi_t)\circ\sigma_t,\chi_{[t+1,\bar t]})-v_{t+1}(s_{t+1},\xi_{[t+1,\bar t]},T_t(\chi'_t)\circ\sigma'_t,\chi'_{[t+1,\bar t]})\mid.
\end{array}\end{equation}

By part (i) of Proposition~\ref{C-onestep}, $\sigma'_t\otimes\chi'_t$ can be made arbitrarily close to $\sigma_t\otimes \chi_t$ by letting $(\sigma'_t,\chi'_t)$ be close enough to $(\sigma_t,\chi_t)$. Then due to Assumption~\ref{f-c}, $M_1$ can be made arbitrarily small by doing the same.

Again, suppose $S=\{\bar s_1,\bar s_2,...\}$. We use the simplified notation that
\begin{equation}
\gamma^{(\prime)}_i(s_t,x_t)=g_t(s_t,x_t,\sigma^{(\prime)}_t\otimes \chi^{(\prime)}_t|\{\bar s_i\}),
\end{equation}
and
\begin{equation}
v_i(\xi_{[t+1,\bar t]})=v_{t+1}(\bar s_i,\xi_{[t+1,\bar t]},T_t(\chi_t)\circ\sigma_t,\chi_{[t+1,\bar t]}).
\end{equation}
Then,~(\ref{m22}) can be expressed as $M_2$ equaling
\begin{equation}\label{m20}
\sup_{(s_t,x_t)\in S\times X,\;\xi_{[t+1,\bar t]}\in(({\cal P}(X))^S)^{\bar t-t}} \mid \sum_i \gamma_i(s_t,x_t)\cdot v_i(\xi_{[t+1,\bar t]})-\sum_i\gamma'_i(s_t,x_t)\cdot v_i(\xi_{[t+1,\bar t]})\mid.
\end{equation}
Let $I(s_t,x_t)$ be the set of $i$'s that induce $\gamma_i(s_t,x_t)\geq \gamma'_i(s_t,x_t)$. Note the $\mid v_i(\xi_{[t+1,\bar t]})\mid$'s are bounded, say by $\overline v$, due to the boundedness of the $\tilde f_{t'}$'s and the finiteness of $\bar t$. Then,~(\ref{m20}) would lead to
\begin{equation}\label{m21}
M_2\leq  2\overline v\cdot \sup_{(s_t,x_t)\in S\times X} \sum_{i\in I(s_t,x_t)} (\gamma_i(s_t,x_t)-\gamma'_i(s_t,x_t)).
\end{equation}
For $\delta$ below $\inf_{s\neq s'}d_S(s,s')$, the event $\rho_S(\tilde g_t(s_t,x_t,\sigma_t\otimes \chi_t),\tilde g_t(s_t,x_t,\sigma'_t\otimes \chi'_t))<\delta$ would trigger
\begin{equation}
\sum_{i\in I(s_t,x_t)} (\gamma_i(s_t,x_t)-\gamma'_i(s_t,x_t))<\delta,
\end{equation}
for every $(s_t,x_t)\in S\times X$; consult~(\ref{consult}) in the proof of Proposition~\ref{V-convergence}. But due to Assumption~\ref{g-c}, the convergence of $\sigma'_t\otimes\chi_t'$ to $\sigma_t\otimes \chi_t$ means that we can make $\tilde g_t(s_t,x_t,\sigma'_t\otimes \chi'_t)$ arbitrarily close to $\tilde g_t(s_t,x_t,\sigma_t\otimes \chi_t)$, at a rate that is independent of $(s_t,x_t)$. Hence, by~(\ref{m21}), $M_2$ can be made arbitrarily small by letting $(\sigma'_t,\chi'_t)$ get close enough to $(\sigma_t,\chi_t)$.

From~(\ref{m33}), we can get
\begin{equation}\begin{array}{l}
M_3\leq\sup_{s_{t+1}\in S,\;\xi_{[t+1,\bar t]}\in(({\cal P}(X))^S)^{\bar t-t}}\mid v_{t+1}(s_{t+1},\xi_{[t+1,\bar t]},T_t(\chi_t)\circ\sigma_t,\chi_{[t+1,\bar t]})\\
\;\;\;\;\;\;\;\;\;\;\;\;\;\;\;\;\;\;\;\;\;\;\;\;\;\;\;\;\;\;\;\;\;\;\;\;-v_{t+1}(s_{t+1},\xi_{[t+1,\bar t]},T_t(\chi'_t)\circ\sigma'_t,\chi'_{[t+1,\bar t]})\mid.
\end{array}\end{equation}
By part (ii) of Proposition~\ref{C-onestep}, $T_t(\chi'_t)\circ \sigma'_t=\sigma'_t\odot\chi'_t\odot\tilde g_t(\cdot,\cdot,\sigma'_t\otimes\chi'_t)$ can be made arbitrarily close to $T_t(\chi_t)\circ \sigma_t=\sigma_t\odot\chi_t\odot\tilde g_t(\cdot,\cdot,\sigma_t\otimes\chi_t)$ by letting $(\sigma'_t,\chi'_t)$ be close enough to $(\sigma_t,\chi_t)$. By the induction hypothesis, $M_3$ can be made arbitrarily small by doing the same.

We have thus completed the induction process. \qed

Here comes the conditional-equilibrium existence result for the transient case.

\begin{theorem}\label{exist1}
	The correspondence $H$ allows for a fixed point $(\sigma_{[2\bar t]},\chi_{[1\bar t]})$, which supplies the game $\Gamma(\sigma_1)$ with an conditional equilibrium $\chi_{[1\bar t]}$.
\end{theorem}

\noindent{\bf Proof of Theorem~\ref{exist1}: }
Due to $S$'s discreteness, ${\cal P}(S)$ is the simplex in $\mathbb{R}^{\mid S\mid}$, whether $\mid S\mid$ be finite or infinite, and hence is compact; the same applies to ${\cal P}(X)$. Thus, ${\cal U}$ is a compact subset of the vector space $\mathbb{R}^{\mid S\mid^{\bar t-1}+\mid X\mid^{\mid S\mid\cdot \bar t}}$, understood as $\mathbb{R}^\infty$ if either $S$ or $X$ is infinite.

For any finite-dimensional $\mathbb{R}^k$, we can take the norm $\mid\mid\cdot\mid\mid$ so that $\mid\mid r\mid\mid=\sum_{l=1}^k \mid r_l\mid/k$ for each $r=(r_l|l=1,...,k)\in\mathbb{R}^k$, whereas for the infinite-dimensional $\mathbb{R}^\infty$, we can let $\mid\mid r\mid\mid=\sum_{l=1}^{+\infty} \mid r_l\mid/2^l$ for each $r=(r_l|l=1,2,...)\in\mathbb{R}^\infty$. A norm thus defined would provide the same convergence as does the weak convergence under Prohorov metric. Since the convex combination of two probabilities is still a probability, ${\cal U}$ is also convex.

For any $(\sigma_{[2\bar t]},\chi_{[1\bar t]})\in {\cal U}$, the set $H(\sigma_{[2\bar t]},\chi_{[1\bar t]})$ is certainly non-empty, for we can construct some $(\sigma'_{[2\bar t]},\chi'_{[1\bar t]})$ belonging to it. First, for $t=2,3,...,\bar t$, we simply let $\sigma'_t=T_{t-1}(\chi_{t-1})\circ\sigma_{t-1}$. Then, for $t=1,2,...,\bar t$ and $s\in S$, let $\chi'_t(s)$ be any measure that assigns its full weight to the set of $x$'s that attain the maximum value $\sup_{y\in X}v_t(s,(\delta_y,\chi_{[t+1,\bar t]}),\sigma_t,\chi_{[t\bar t]})$.

Now we show that $H^S:{\cal U}\Rightarrow {\cal S}$ and $H^X:{\cal U}\Rightarrow {\cal X}$ are closed- and convex-valued, as well as upper hemi-continuous. These would lead to the same properties for $H$. According to~(\ref{hs-def}), each $H^S(\sigma_{[2\bar t]},\chi_{[1\bar t]})$ contains exactly one point, and hence is automatically closed and convex. For the upper hemi-continuity property, we need only to show that the value contained in $H^S(\sigma_{[2\bar t]},\chi_{[1\bar t]})$ moves continuously with both $\sigma_{[2\bar t]}$ and $\chi_{[1\bar t]}$. But this has been guaranteed by part (ii) of Proposition~\ref{C-onestep}.

According to~(\ref{hx-def}), each $H^X(\sigma_{[2\bar t]},\chi_{[1\bar t]})$ is a set of probability vectors, with each component probability assigning the full measure to a particular measurable set. This set of probability vectors is certainly convex. To show that it is closed, suppose $\chi'_{m,[1\bar t]}$ for $m=1,2,...$ form a sequence in $H^X(\sigma_{[2\bar t]},\chi_{[1\bar t]})$ that converges to a given $\chi'_{[1\bar t]}$. We are to show that
\begin{equation}\label{above0}
\chi'_{[1\bar t]}\in H^X(\sigma_{[2\bar t]},\chi_{[1\bar t]}).
\end{equation}
Now for any $t=1,2,...,\bar t$, $s\in S$, and $\epsilon>0$, as long as $m$ is large enough,
\begin{equation}
\chi'_t(s|(\tilde X_t(s,\sigma_{[2\bar t]},\chi_{[1\bar t]}))^\epsilon)\geq \chi'_{mt}(s|\tilde X_t(s,\sigma_{[2\bar t]},\chi_{[1\bar t]}))-\epsilon=1-\epsilon.
\end{equation}
Due to 
the arbitrariness of $\epsilon$, this means $\chi'_t(s|\tilde X_t(s,\sigma_{[2\bar t]},\chi_{[1\bar t]}))=1$, and hence~(\ref{above0}) is true.

We now show that $H^X$ is upper hemi-continuous. Let $\sigma_{m,[2\bar t]}$ be a sequence in ${\cal S}$ that converges to a given $\sigma_{[2\bar t]}$, $\chi_{m,[1\bar t]}$ a sequence in ${\cal X}$ that converges to a given $\chi_{[1\bar t]}$, and $\chi'_{m,[1\bar t]}$ another sequence in ${\cal X}$ that converges to a given $\chi'_{[1\bar t]}$. Suppose for each $m=1,2,...$,
\begin{equation}\label{hyp}
\chi'_{m,[1\bar t]}\in H^X(\sigma_{m,[2\bar t]},\chi_{m,[1\bar t]}),
\end{equation}
we are to show that
\begin{equation}\label{target}
\chi'_{[1\bar t]}\in H^X(\sigma_{[2\bar t]},\chi_{[1\bar t]}).
\end{equation}
By~(\ref{hx-def}), we see that~(\ref{hyp}) for each $m$ indicates that, for each $t=1,2,...,\bar t$ and $s\in S$,
\begin{equation}\label{hyp0}
\chi'_{mt}(s|\tilde X_t(s,\sigma_{mt},\chi_{m,[t\bar t]}))=1;
\end{equation}
whereas,~(\ref{target}) boils down to that, for each $t=1,2,...,\bar t$ and $s\in S$,
\begin{equation}\label{target0}
\chi'_t(s|\tilde X_t(s,\sigma_t,\chi_{[t\bar t]}))=1.
\end{equation}

We fix some $t$ and $s$. Let $\epsilon>0$ be small enough, so that 
there is no need to distinguish between $(X')^\epsilon$ and $X'$ for any $X'\subseteq X$. Now since $\chi'_{mt}$ converges to $\chi'_t$, for $m$ large enough,
\begin{equation}
\chi'_t(s|\tilde X_t(s,\sigma_t,\chi_{[t\bar t]}))=\chi'_t(s|(\tilde X_t(s,\sigma_t,\chi_{[t\bar t]}))^\epsilon)\geq \chi'_{mt}(s|\tilde X_t(s,\sigma_t,\chi_{[t\bar t]}))-\epsilon.
\end{equation}
For the time being, suppose $\tilde X_t(s,\sigma_t,\chi_{[t\bar t]})$ is known to be upper hemi-continuous in $(\sigma_t,\chi_{[t\bar t]})$.
By also noting the hypothesis on the convergence of $\sigma_{mt}$ to $\sigma_t$ and that of $\chi_{m,[t\bar t]}$ to $\chi_{[t\bar t]}$, we can obtain, for $m$ large enough,
\begin{equation}
\tilde X_t(s,\sigma_{mt},\chi_{m,[t\bar t]})\subseteq (\tilde X_t(s,\sigma_t,\chi_{[t\bar t]}))^\epsilon=\tilde X_t(s,\sigma_t,\chi_{[t\bar t]}).
\end{equation}
Thus, for $m$ large enough,
\begin{equation}\label{lyp}
\chi'_t(s|\tilde X_t(s,\sigma_t,\chi_{[t\bar t]}))\geq \chi'_{mt}(s|\tilde X_t(s,\sigma_t,\chi_{[t\bar t]}))-\epsilon\geq \chi'_{mt}(s|\tilde X_t(s,\sigma_{mt},\chi_{m,[t\bar t]}))-\epsilon,
\end{equation}
which, according to~(\ref{hyp0}), is above $1-\epsilon$. In view of the arbitrariness of $\epsilon$, we can achieve~(\ref{target0}).

We now come back to the upper hemi-continuity of $\tilde X_t(s,\cdot)$ as a correspondence from ${\cal P}(S)\times (({\cal P}(X))^S)^{\bar t-t+1}$ to $X$. Suppose $\sigma_{mt}$ converges to $\sigma_t$, $\chi_{m,[t\bar t]}$ converges to $\chi_{[t\bar t]}$, and $x_m$ converges to $x$. For every $m=1,2,...$, suppose $x_m\in \tilde X_t(s,\sigma_{mt},\chi_{m,[t\bar t]})$, which, by~(\ref{xx-def}), means
\begin{equation}\label{arg1}
v_t(s,(\delta_{x_m},\chi_{m,[t+1,\bar t]}),\sigma_{mt},\chi_{m,[t\bar t]})\geq v_t(s,(\delta_y,\chi_{m,[t+1,\bar t]}),\sigma_{mt},\chi_{m,[t\bar t]}),\hspace*{.5in}\forall y\in X.
\end{equation}
By $X$'s discreteness, $x_m$ would be $x$ for sufficiently large $m$. This, combined with Proposition~\ref{V-continuity} and the hypothesis on the convergence of $\sigma_{mt}$ to $\sigma_t$ and that of $\chi_{m,[t\bar t]}$ to $\chi_{[t\bar t]}$, would entail that, for any $\epsilon>0$, as long as $m$ is large enough,
\begin{equation}\label{arg2}\begin{array}{l}
v_t(s,(\delta_x,\chi_{[t+1,\bar t]}),\sigma_t,\chi_{[t\bar t]})\geq v_t(s,(\delta_x,\chi_{m,[t+1,\bar t]}),\sigma_{mt},\chi_{m,[t\bar t]})-\epsilon\\
\;\;\;\;\;\;\;\;\;\;\;\;=v_t(s,(\delta_{x_m},\chi_{m,[t+1,\bar t]}),\sigma_{mt},\chi_{m,[t\bar t]})-\epsilon\\
\;\;\;\;\;\;\;\;\;\;\;\;\geq v_t(s,(\delta_y,\chi_{m,[t+1,\bar t]}),\sigma_{mt},\chi_{m,[t\bar t]})-\epsilon \geq v_t(s,(\delta_y,\chi_{[t+1,\bar t]}),\sigma_t,\chi_{[t\bar t]})-2\epsilon,
\end{array}\end{equation}
for any $y\in X$. Since $\epsilon$ can be arbitrarily small, we see from~(\ref{xx-def}) that $x\in\tilde X_t(s,\sigma_t,\chi_{[t\bar t]})$. Thus we have the upper hemi-continuity of $\tilde X_t(s,\cdot)$.

In summary, $H$ is a non-empty, closed- and convex-valued, as well as upper hemi-continuous correspondence on the compact and convex subset ${\cal U}$ that is embedded in a normed linear topological space. We can therefore apply the Kakutani-Fan-Glicksberg fixed point theorem to verify that $H$ has a fixed point. \qed

\subsection{The Stationary Case}\label{app-f-s}

Denote the space ${\cal P}(S)$ by ${\cal S}_\infty$ and the space $({\cal P}(X))^S$ by ${\cal X}_\infty$. Let ${\cal U}_\infty={\cal S}_\infty\times {\cal X}_\infty$. Define a correspondence $H_\infty:{\cal U}_\infty\Rightarrow {\cal U}_\infty$, so that for any $\sigma\in {\cal S}_\infty$ and $\chi\in {\cal X}_\infty$,
\begin{equation}
H_\infty(\sigma,\chi)=H^S_\infty(\sigma,\chi)\times H^X_\infty(\sigma,\chi),
\end{equation}
where
\begin{equation}\label{hs-def-s}
H^S_\infty(\sigma,\chi)=\{\sigma'\in {\cal S}_\infty|\sigma'=T(\chi)\circ \sigma\},
\end{equation}
and
\begin{equation}\label{hx-def-s}
H^X_\infty(\sigma,\chi)=\{\chi'\in {\cal X}_\infty|\chi'(s|\tilde X_\infty(s,\sigma,\chi))=1,\;\forall s\in S\}.
\end{equation}

A fixed point $(\sigma,\chi)$ for $H_\infty$ would provide a stationary Markov equilibrium $\chi$ for the stationary nonatomic game $\Gamma^\infty$ in the sense of~(\ref{first-alt-s}), with $\sigma$ supplying the invariant deterministic environment that is generated from all players adopting policy $\chi$. To show that such an equilibrium exists, we first need the following consequence of Proposition~\ref{V-continuity}.

\begin{proposition}\label{V-continuity-s}
	The value $v^\infty(s,\xi_{[1\infty]},\sigma,\chi^\infty)$ defined in~(\ref{recursive-s}) is continuous in $\sigma\in {\cal S}_\infty$ and $\chi\in {\cal X}_\infty$ at an $(s,\xi_{[1\infty]})$-independent rate.
\end{proposition}
\noindent{\bf Proof of Proposition~\ref{V-continuity-s}: }From~(\ref{bounded-s}), we see that
\begin{equation}\label{d3}
\mid v^\infty(s,\xi_{[1\infty]},\sigma,\chi^\infty)-v^t(s,\xi_{[1t]},\sigma,\chi^t)\mid\leq \frac{\bar\alpha^t\cdot\overline f}{1-\bar\alpha}.
\end{equation}
Thus, for any $\epsilon>0$, by fixing at a large enough $t$, we can ensure
\begin{equation}\label{need1}
\mid v^\infty(s,\xi_{[1\infty]},\sigma'',(\chi'')^\infty)-v^t(s,\xi_{[1t]},\sigma'',(\chi'')^t)\mid<\frac{\epsilon}{3},
\end{equation}
for any $s$, $\xi_{[1\infty]}$, $\sigma''$, and $\chi''$. At the same time, Proposition~\ref{V-continuity} means that, for $(\sigma',\chi')$ close enough to any given $(\sigma,\chi)$, we can guarantee
\begin{equation}\label{need2}
\mid v^t(s,\xi_{[1t]},\sigma,\chi^t)-v^t(s,\xi_{[1t]},\sigma',(\chi')^t)\mid<\frac{\epsilon}{3},
\end{equation}
for any $s$ and $\xi_{[1t]}$. Then,
\begin{equation}\begin{array}{l}
\mid v^\infty(s,\xi_{[1\infty]},\sigma,\chi^\infty)-v^\infty(s,\xi_{[1\infty]},\sigma',(\chi')^\infty)\mid\leq \mid v^\infty(s,\xi_{[1\infty]},\sigma,\chi^\infty)\\
\;\;\;\;\;\;\;\;\;\;\;\;\;\;\;\;\;\;\;\;\;\;\;\;-v^t(s,\xi_{[1t]},\sigma,\chi^t)\mid+\mid v^t(s,\xi_{[1t]},\sigma,\chi^t)-v^t(s,\xi_{[1t]},\sigma',(\chi')^t)\mid\\
\;\;\;\;\;\;\;\;\;\;\;\;\;\;\;\;\;\;+\mid v^t(s,\xi_{[1t]},\sigma',(\chi')^t)-v^\infty(s,\xi_{[1\infty]},\sigma',(\chi')^\infty)\mid<\epsilon.
\end{array}\end{equation}
Thus, $v^\infty(s,\xi_{[1\infty]},\sigma,\chi^\infty)$ is continuous in $(\sigma,\chi)$ at an $(s,\xi_{[1\infty]})$-independent rate. \qed

We can then have the desired conditional-equilibrium existence result by using the Kakutani-Fan-Glicksberg fixed point theorem.

\begin{theorem}\label{exist1-s}
	The correspondence $H_\infty$ allows for a fixed point $(\sigma,\chi)$, which supplies the game $\Gamma^\infty$ with an equilibrium $\chi$.
\end{theorem}
\noindent{\bf Proof of Theorem~\ref{exist1-s}: }Due to the discreteness of $S$ and $X$, ${\cal U}_\infty$ is a compact subset of the vector space $\mathbb{R}^{\mid S\mid+\mid X\mid^{\mid S\mid}}$, understood as $\mathbb{R}^\infty$ if either $S$ or $X$ is infinite. Regardless of whether the space is finite- or infinite-dimensional, we can take the norm adopted in the proof of Theorem~\ref{exist1}. Since the convex combination of two probabilities is still a probability, ${\cal U}_\infty$ is convex.

Using virtually the same corresponding arguments in the proof of Theorem~\ref{exist1}, we can show that $H_\infty(\sigma,\chi)$ at any $(\sigma,\chi)\in {\cal U}_\infty$ is non-empty, closed, and convex. We separate the upper hemi-continuity of $H_\infty$ into that for $H^S_\infty$ and that for $H^X_\infty$.

The upper hemi-continuity of $H^S_\infty$ again comes from Proposition~\ref{C-onestep}. Furthermore, we can use almost the same arguments from~(\ref{arg1}) to~(\ref{arg2}), this time relying on Proposition~\ref{V-continuity-s} instead of Proposition~\ref{V-continuity}, to show that $\tilde X_\infty(s,\cdot)$ as a correspondence from ${\cal P}(S)\times ({\cal P}(X))^S$ to $X$ is upper hemi-continuous. Then, using almost the same arguments from~(\ref{hyp}) to~(\ref{lyp}), we can verify that $H^X_\infty$ is upper hemi-continuous.

With all these properties, we can apply the Kakutani-Fan-Glicksberg fixed point theorem to verify that $H_\infty$ has a fixed point. \qed

\end{document}